\documentclass[12pt]{article}
\usepackage[margin=0.95 in]{geometry}
\usepackage{amsmath} 
\usepackage{amssymb,amsfonts}
\usepackage[all]{xy}
\usepackage{graphicx}

\usepackage{amsmath}
\usepackage{amssymb}
\usepackage{float}
\usepackage{array}
\usepackage{tikz}
\usepackage{mathtools}
\usepackage{mathrsfs} 
\usepackage[hidelinks]{hyperref}
\usepackage{cite}
\usepackage{bbm}
\numberwithin{equation}{section}
\usepackage[utf8x]{inputenc}

\newcommand{\be}{\begin{equation}}
\newcommand{\ee}{\end{equation}}
\def\bea{\begin{eqnarray}}
\def\eea{\end{eqnarray}}

\numberwithin{equation}{section}
\numberwithin{table}{section}

\begin{document}

\hypersetup{pageanchor=false}
\begin{titlepage}
\vbox{
    \halign{#\hfil         \cr
           } % end of \halign
      }  % end of \vbox
\vspace*{15mm}
\begin{center}
{\Large \bf 
Inverse Monoid Topological Quantum Field Theories}

 \vspace*{3mm}

 {\Large \bf and }

\vspace*{3mm}

{\Large \bf Open-Closed  Grand Canonical  Symmetric Orbifolds
}

\vspace*{15mm}

{\large  Jan Troost }
\vspace*{8mm}

%$^b$ 
Laboratoire de Physique de l'\'Ecole Normale Sup\'erieure \\ 
 \hskip -.05cm
 CNRS, ENS, Universit\'e PSL,  Sorbonne Universit\'e, \\
 Universit\'e de Paris 
 \hskip -.05cm F-75005 Paris, France	 
\vskip 0.8cm
	{\small
		E-mail:
		\texttt{ jan.troost@ens.fr}
	}
\vspace*{0.8cm}
\end{center}

\begin{abstract}
We present an open-closed topological quantum field theory for  inverse monoids which generalizes the theory of principle fiber bundles with finite gauge group over Riemann surfaces with boundary. The theory is constructed using the isomorphism between the semisimple inverse monoid algebra and a matrix algebra which lies at the heart of monoid structure and representation theory. An example that we study in detail  is the Ivanov-Kerov monoid of partial permutations. We review motivations from string theory for the resulting grand canonical theory of covers with boundaries. 
\end{abstract}

\end{titlepage}
\hypersetup{pageanchor=true}

\tableofcontents

\section{Introduction}
  To all finite groups $G$  there is a canonically associated open-closed topological quantum field theory in two dimensions \cite{Atiyah:1989vu,Dijkgraaf:1989pz,Fukuma:1993hy}. The center of the group algebra 
  %$\mathbb{C}[G]$
  is the algebra of bulk observables and the representations of the group are the set of boundary conditions. 
  There are specific examples of these theories that are of particular interest. An instructive example is the theory associated to the symmetric group $S_n$ of  permutations of a set of $n$ elements. One of its important roles is that it coincides with the Hurwitz theory that counts covers of Riemann surfaces of degree $n$ \cite{Hurwitz} or principal bundles over the Riemann surface with gauge group $S_n$ \cite{Mednyh}. To the extent that counting problems involving interchangeable objects are universal in physics, the topological theories that code these counting problems are as well. 

It is natural to attempt to construct a generalization of these topological quantum field theories that leaves the  number $n$ of interchangeable objects arbitrary. Indeed, this is akin to  developing the grand canonical perspective in statistical physics. In the context of string theories,  the argument for generalization boils down to the observation that in string theory one defines observables in terms of a sum over covers of all possible degrees. This general argument has  incarnations in the AdS$_3$/CFT$_2$ duality \cite{Maldacena:2000hw,Giveon:1998ns,Eberhardt:2020bgq}, in the Hurwitz/Gromov-Witten correspondence \cite{Okounkov:2002cja,Benizri:2024mpx} and in the application of symmetric group gauge theory to two-dimensional Yang-Mills theory via Schur-Weyl duality \cite{Gross:1993hu,Cordes:1994fc,Benizri:2025xmz}. 
In this paper, we construct such a grand canonical theory of arbitrary degree covers with boundary. 

The mathematical structure that replaces the role of the group $S_n$ of permutations of sheets of the cover is the Ivanov-Kerov finite inverse monoid\footnote{An inverse monoid $M$ has an associative multiplication with an identity and for each element $m \in M$ an element $m^\ast$ such that $m \, m^\ast \, m = m$ and $m^\ast \, m \, m^\ast =m^\ast$. The monoid is finite if it has a finite number of elements.} of partial permutations \cite{IK,Benizri:2024mpx} in which one specifies the sheets (including  their number) on which the permutation group acts. Thus, we formulate the intermediate as well as the broader goal to generalize the open-closed topological field theory to finite inverse monoids. We realize this objective by exploiting the structure theory of monoids. Every finite inverse monoid algebra is isomorphic to a matrix algebra -- see \cite{Steinberg} and references therein -- and this allows to systematically formulate a corresponding  open-closed topological quantum field theory \cite{Lauda:2006mn}.  We review these facts at the hand of concrete formulas which simplify, exemplify and generalize the  mathematical treatment of the subject. 

Let us remark briefly on how our analysis relates to various strains in the literature.
At fixed degree $n$, there is a close connection between the open-closed theory with gauge group $S_n$ and  D-branes in symmetric orbifolds (see e.g. \cite{Belin:2021nck,Gaberdiel:2021kkp}). 
There is also a relation to the theory of seamed surfaces \cite{AN}. 
These studies concentrate on fixed degree and/or particular boundary conditions. 
In the limit where we allow for an arbitrarily large degree $n$, we recall that the closed Ivanov-Kerov gauge theory or 
closed Hurwitz theory corresponds to Gromov-Witten theory on Riemann surfaces \cite{Okounkov:2002cja}. This is a gauge/gravity correspondence that is a topologically twisted version of an AdS/CFT correspondence \cite{Gopakumar:2011ev,Li:2020zwo,Benizri:2024mpx}. Given the open-closed grand canonical Hurwitz theory constructed here, we must expect the theory to have a gravity dual that also includes boundaries. There are some theories of open surfaces on $\mathbb{CP}^1$ \cite{Elitzur:2011ug,Buryak:2020pgl}. These theories concentrate on one particular choice of boundary condition.

The plan of the paper is as follows. In section \ref{DWGroup} we recall the open-closed topological quantum field theory for the group case from various complementary perspectives. The example  serves as a concrete handle for the more abstract sections to follow and may be a helpful unifying review of the literature.  Section \ref{FiniteInverseMonoids} firstly recalls  the structure theory for finite inverse monoids and secondly constructs an open-closed topological field theory for inverse monoids.   
We provide an application of the general theory in section \ref{GrandCanonicalOpenClosed} where we specify the theory of section \ref{FiniteInverseMonoids} to the Ivanov-Kerov monoid of partial permutations and present the grand canonical theory of open and closed covering surfaces with generic boundary conditions. We draw conclusions and broaden the discussion in section \ref{Conclusions}.
Appendix \ref{IdempotencyAppendix} contains details on a technical heart of the paper. 

\section{The Finite Group Topological Quantum Field Theory}
\label{DWGroup}
\label{GroupTQFT}
We review the open-closed two-dimensional topological field theory for a finite gauge group $G$ \cite{Atiyah:1989vu,Dijkgraaf:1989pz,Fukuma:1993hy}. The theory will serve as the basic model that we will extend into new territory. 
We discuss it in a practical language in terms of group theoretic properties, but also in  more abstract  terms in order to prepare for the sections to come.  
This section is also based on \cite{Lauda:2006mn,Moore:2006dw} and the  notes \cite{Moore,Pennig}.

\subsection{The Bulk Theory}
We review the bulk operator algebra for the open-closed two-dimensional topological quantum field theory associated to a finite group $G$. Each finite group has a corresponding group algebra $\mathbb{C}[G]$ consisting of linear combinations of group elements with complex coefficients.\footnote{We work over the field of complex numbers which has characteristic zero. When the characteristic of the field divides the order of the finite group, the mathematics of the group algebra is more intricate.}
The center $Z(\mathbb{C}[G])$ of the group algebra is the set of  conjugation invariant linear combinations of the group elements. A basis of such combinations is given by the sum $C_{[g]}$ over group elements $g$ in each conjugacy class $[g]$:
\begin{equation}
C_{[g]} = \sum_{g \in [g]} g \, .
\end{equation} 
The basis has a number of elements equal to the number $r$ of conjugacy classes.
An important property of the finite group algebra $\mathbb{C}[G]$ is that it is semisimple.
It is therefore a product of matrix algebras.\footnote{By the Wedderburn-Artin theorem, a semisimple algebra over the complex numbers is  a product of matrix algebras over division algebras of $\mathbb{C}$. The only division algebra of $\mathbb{C}$ is $\mathbb{C}$ itself.} The center of the algebra can therefore be modeled by $\mathbb{C}^r=Z(\mathbb{C}[G])$.
There are orthogonal idempotents in the center that project onto the simple factors $\mathbb{C}$ of the center group algebra. For the finite group algebra, they are equal to 
\begin{equation}
e_\rho = \sum_{g \in G} \frac{d_\rho}{|G|} \chi_\rho(g^{-1}) g \, ,
\label{GroupIdempotent}
\end{equation}
where $\rho$ labels the set $\hat{G}$ of irreducible representations $V_\rho$ of the finite group $G$, the characters $\chi_\rho$ are class functions, $|G|$ is the cardinality of the group and $d_\rho$ denotes the dimension of the representation $V_\rho$.\footnote{We include a proof of the idempotency of these linear combinations in appendix \ref{IdempotencyAppendix}.}
The topological quantum field theory has correlators determined by the group algebra
and the bulk one-point function:
\begin{equation}
\langle g \rangle_{\text{bulk}} =  \frac{1}{|G|} \delta_{1,g} \, .
\end{equation}
For instance, the partition function $Z_g$ can be computed on any closed Riemann surface of genus $g$ and it equals\footnote{We pick a particular and simple bulk theory in the rest of the section. We will come back to the choices implicitly made here at various junctions of our exposition.}
\begin{equation}
Z_g = \sum_{\rho} \left( \frac{d_\rho}{|G|} \right)^{2-2g} \, .
\end{equation}
 The one-point functions evaluated on the idempotents (\ref{GroupIdempotent}) are an invariant of the Frobenius algebra equal to:
\begin{equation}
\langle e_\rho \rangle_{\text{bulk}} = \frac{d_\rho^2}{|G|^2} \, .
\end{equation}
Up to a factor of $1/|G|$, the one-point function of the central idempotents equals the Plancherel measure on the space of irreducible representations of the group. 
The calculation of the structure constants of the algebra in the conjugacy class sum basis, which is equivalent to the determination of the three-point functions in the topological quantum field theory, is an interesting branch of finite group theory and combinatorics and finds wide application in physics.\footnote{For an application to the $AdS_3/CFT_2$ duality in this spirit, see \cite{Li:2020zwo}.}

\subsection{The Theory with Boundaries}
The open-closed theory on Riemann surfaces with boundary has been presented in multiple interesting ways \cite{Moore:2006dw,Moore,Lauda:2006mn}. We revisit these perspectives and clarify how one point of view suggests a generalization of another and vice versa.  We  first discuss the theory
from an abstract semisimple perspective, then  more concretely  in terms of explicit group theory expressions (which may be closest to its various applications in the physics literature) \cite{Moore:2006dw,Moore} and then in terms of matrix algebras \cite{Lauda:2006mn}. We provide a fourth perspective that further indicates how various natural choices can be made among the zoo of theories. The full treatment is in the spirit of preparing for the more intricate construction of the inverse monoid theory in the next section. We do assume that the reader is familiar with a number of abstract properties and axioms of open-closed topological quantum field theory. See e.g. \cite{Moore:2006dw,Moore,Lauda:2006mn,Lazaroiu:2000rk,Koch} for background.

\subsubsection{The Panoramic Semisimple  View}
\label{SemisimplePerspective}
The group algebra $\mathbb{C}[G]$ is semisimple.  The bulk operator algebra $Z(\mathbb{C}[G])$ is therefore equivalent to $\mathbb{C}^r$ and we can apply our knowledge of the general structure of the open theory corresponding to a closed theory with a semisimple bulk operator algebra \cite{Moore:2006dw,Moore}. 
If the commutative bulk operator algebra $\mathcal{C}=\mathbb{C}^r$ is semisimple and of dimension $r$, then its spectrum $X=\text{spec}(\mathcal{C})$ can be considered to be the space-time of the topological theory. For the algebra at hand, the spectrum is equal to $r$ points.  The number of points equals   the number of conjugacy classes of the group.
The category of boundary conditions ${\cal B}$ becomes equivalent to the category of vector bundles on $X$.
The vector bundles over a discrete set of points  are an assignment of vector spaces to each of those points. We have a vector space $V_\lambda$ for each point $\lambda \in X=\text{spec}(\mathbb{C}^r)$.
The choice of vector space is merely a choice of dimension
 $m_\lambda$ for each point $\lambda$ in the spectrum of the bulk algebra. The boundary operators are linear maps between the vector spaces associated to the boundary conditions to its left and right. Up to a choice of linear form (which depends on $r$ parameters), we have described the full theory abstractly.  This is an interesting and powerful perspective to keep in mind. However, the perspective provides little intuition for choosing  parameters such that the topological theory corresponds to one with a nice geometric or algebraic interpretation.

\subsubsection{The Group Representation Theoretic View}
\label{GroupPerspective}
A second perspective is to think of the set of boundary conditions ${\cal B}$ as the set of representations $R=m_\lambda \lambda$ of the group $G$ \cite{Moore:2006dw,Moore}. The numbers $m_\lambda$ are the positive multiplicities of the irreducible representations $\lambda$ in the decomposition of the representation $R$.
The boundary operators, residing between two boundaries with boundary labels $R_1$ and $R_2$, are the group compatible homomorphisms between the representations \cite{Moore:2006dw,Moore}. Thus, they are equivalent to matrix spaces $\prod_{i=1}^r {M_{m_1^\lambda \times m_2^\lambda}} (\mathbb{C})$. The boundary operators between equal boundary conditions are square matrix algebras. The boundary one-point functions for such operators are proportional to the traces of the matrices. This  concrete view on the open-closed theory comes with a set of definitions of the basic operations which include the boundary and the bulk trace as well as adjoint maps from the bulk to the boundary operator algebras and back. They need to satisfy consistency ("sewing") conditions \cite{Lazaroiu:2000rk,Moore:2006dw} which includes a topological incarnation of the Cardy condition. For $\psi$ a boundary operator associated to a representation $R$ and $c_g g$ (summed over $g$) a bulk operator, the definitions are as follows. Firstly, one chooses a boundary trace normalized as \cite{Moore:2006dw,Moore}:
\begin{align}
\langle \psi \rangle_{\text{boundary}} &= \frac{1}{|G|} \text{Tr}_V \psi \,  \label{BoundaryTrace}
\end{align}
where $V$ is the representation space of the representation $R$. 
The map $i$ from bulk to the boundary operators is:
\begin{align}
i (c_g g) &= c_g R(g) \label{Mapi}
\end{align}
where $c_g g$ lies in the center of the group algebra.
The adjoint map is:
\begin{align}
i^\ast (\psi) &= \sum_\lambda \text{Tr} (P_\lambda \psi) \chi_\lambda (g^{-1}) g \, , \label{MapiStar}
\end{align}
where $P_\lambda \psi$ is the projection on the component of the homomorphism $\psi$ associated to the irreducible representation $\lambda$ and the trace is over the multiple $\lambda$ representations only. 
We will check the consistency conditions on these definitions shortly. 

\subsubsection{The Matrix Algebras}
\label{MatrixPerspective}
A third perspective we provide is in the context of the state sum construction of  open-closed topological quantum field theories \cite{Lauda:2006mn}\footnote{Concentrating on theories with a state sum construction is akin to studying theories that allow for a lattice description. We impose this requirement, thus excluding a class of theories. See \cite{Lauda:2006mn}.}, in particular for matrix algebras. It occupies an intermediate role between the panoramic and the group theoretic perspective.

We recall that the homomorphisms from a representation $R$ to itself can be represented as square matrices with dimensions equal to the multiplicities of the representations. It will be convenient to remind ourselves that these homomorphisms are always equal to the identity in the attached irreducible representation spaces and we will therefore accompany these matrices with an identity matrix in the representation. For multiplicities $m_\lambda$, we therefore consider the boundary algebras -- compare to \cite{Lauda:2006mn} --:
\begin{align}
A &= \oplus_{\lambda} M_{m_\lambda}(\mathbb{C}) \otimes \{ 1_{d_\lambda \times d_\lambda} \}\, . \label{BoundaryMatrixAlgebra}
\end{align}
For simplicity we will assume that the multiplicities $m_\lambda$ are non-zero for all $\lambda$.\footnote{If they are not, certain bulk operators decouple from the open theory. The open-closed theory can still be constructed in essentially the same manner.}
The idempotent and central elements in the boundary algebra are:
\begin{align}
z_\lambda &= \sum_{p=1}^{m_\lambda} e_{pp}^{(\lambda)} \, ,
\end{align}
where $e_{pq}^{(\lambda)}$ are matrices with unit at the $(p,q)$ entry (in the $\lambda$ summand) and zero otherwise.  The trailing unit matrix is implicit. 
We pick a boundary trace which implies a choice of normalization for each simple summand in the algebra:
\begin{align}
\langle e_{pq}^{(\lambda)} \rangle_{\text{boundary}} &= \delta_{pq} \, \alpha_\lambda \, . \label{AlphaParameters}
\end{align}
In fact, by state sum reconstruction \cite{Lauda:2006mn}, the choice of parameters $\alpha_\lambda$ is equivalent to a choice of bulk trace parameters.
%\footnote{Given the universality of the bulk sector, it may more intuitive to think of these parameters as specifying the bulk theory and implying a boundary trace.}
With the linear form and the multiplicities in hand, one can compute the window element $a$  \cite{Lauda:2006mn} that determines the weight that punctures contribute in the partition function:
\begin{align}
a &= \sum_{\lambda} \frac{m_\lambda}{\alpha_\lambda} z_\lambda \, .
\end{align}
Given the linear form, the window element is fixed by the multiplicities $m_\lambda$. 
These data determine an open-closed theory fully -- see \cite{Lauda:2006mn} for all the details. For instance, the partition function for a surface of genus $g$ with $k$ windows or punctures equals:
\begin{equation}
Z_{g,k} = \sum_\lambda (\frac{m_\lambda}{\alpha_\lambda})^k \alpha_\lambda^{-2(g-1)} \, .
\end{equation}

\subsubsection{The Consistency and Choices Made}
At this point, we have the tools in hand to both discuss the consistency conditions on the finite group theory with boundaries and the specific choices that are made to obtain the theory when we consider it from a more generic semisimple or matrix algebra perspective.

There are various manners in which one may check consistency and they have been detailed in the literature to a varying degree \cite{Moore:2006dw,Moore,Lauda:2006mn}. We verify the consistency conditions once more and link up the various perspectives we laid out. Firstly, the map $i$ in equation (\ref{Mapi}) is a homomorphism  because $R$ is a representation. Secondly, the map $i$ sends the bulk unity operator to the boundary unit operator - this normalizes the map. 
The image of the map $i$ is central in the boundary algebra because of the definition of the boundary operators as group compatible homomorphisms. 
Importantly, the theory discussed in\cite{Moore:2006dw,Moore} singles out the choice of one-point function for boundary operators  which is:
\begin{align}
\langle \psi \rangle_{\text{boundary}} &= \frac{1}{|G|} \text{Tr}_V \psi
\end{align}
where $V$ is the representation space of the representation $R$.
In the matrix algebra perspective (\ref{AlphaParameters}), this singles out the parameter choice  $\alpha_\lambda=d_\lambda/|G|$. The factor of $d_\lambda$ arises from the unit matrix explained around equation (\ref{BoundaryMatrixAlgebra}).
Given these facts, we can compute which operator corresponds to the central operator $i(c_g g)$ by exploiting  Schur's lemma and taking traces. We find that:
\begin{align}
i(c_g g) &= \sum_{\lambda,g} c_g \frac{\chi_\lambda(g)}{d_\lambda}1_{m_\lambda \times m_\lambda} \otimes 1_{d_\lambda \times d_\lambda} \, .
\end{align}
At this stage, we will leave a few parameters free to demonstrate how consistency conditions fix them. 
%We've chosen to think of our bulk operators as in the center of $\mathbb{C}[G]$. 
In the bulk, we have a trace proportional to $\langle g \rangle_{\text{bulk}}=N \delta_{1,g}$.
Given this inner product, we can compute the adjoint operator $i^\ast$. We have:
\begin{align}
\langle i(c_g g) \psi \rangle_{\text{boundary}} 
&= \frac{1}{|G|} \sum_{g,\lambda} c_g \frac{\chi_\lambda(g)}{d_\lambda} \text{Tr}_{\lambda} (P_\lambda \psi) \, 
\end{align}
and must compare  to the adjoint ansatz -- we added a free parameter to equation (\ref{MapiStar}) for illustration purposes --:
\begin{align}
i^\ast (\psi) &= M \sum_\lambda \text{Tr} (P_\lambda \psi) \frac{\chi_\lambda (h^{-1})}{d_\lambda} h
\end{align}
and therefore to
\begin{align}
\langle i^\ast(\psi) c_g g \rangle_{\text{bulk}} &= MN c_g \sum_\lambda tr (P_\lambda \psi)\frac{ \chi_\lambda (g)}{d_\lambda} \, .
\end{align}
The operators $i$ and $i^\ast$ are adjoint provided that 
 $M=(N |G|)^{-1}$. 
Last but not least, we need to impose the Cardy condition. It implies that the duality map ${\pi_2}^1$ from the boundary operators with boundary conditions $R_1$ to those with boundary conditions $R_2$ equals the composition of adjoints $i_2$ and $i^{1 \ast}$ \cite{Moore:2006dw,Moore}.
We must have:
\begin{align}
{\pi_2}^1(\psi) &= (i_2 \circ i^{1 \ast}) (\psi) \label{CardyCondition}
\end{align} where the left-hand side is defined by a sum 
\begin{equation}
{\pi_2}^1(\psi) =\psi_\alpha \psi \psi^\alpha 
\end{equation}
over dual bases $\psi_\alpha$ $(\psi^\alpha)$ of the boundary operators with boundary conditions $R_{1(2)}$ and $R_{2(1)}$. The right-hand side of the condition (\ref{CardyCondition}) evaluates to:
\begin{align}
(i_2 \circ i^{1 \ast}) (\psi)
%&= i(M tr (P_\mu \psi) \frac{\chi_\mu(g^{-1})}{d_\mu} g)
%\nonumber \\
&= M \sum_{g,\mu,\nu} \text{Tr} (P_\mu \psi) \frac{\chi_\mu(g^{-1})}{d_\mu} \frac{\chi_\nu(g)}{d_\nu} 1_{R_2} \otimes 1_{d_\nu \times d_\nu}
\nonumber \\
%&= M tr (P_\mu \psi) \frac{\chi_\mu(g^{-1})}{d_\mu} \frac{\chi_\nu(g)}{d_\nu} 1_{R_2} \otimes 1_{d_\nu\times d_\nu}
%\nonumber \\
&= M \frac{|G| }{d_\mu^2} \text{Tr} (P_\mu \psi) 1_{R_2} \otimes 1_{d_\mu\times d_\mu}
\, .
\end{align}
To determine the normalization $M$, we recall that for   $\psi_\alpha$ and $\psi^\alpha$ dual bases for $\langle \cdot \rangle_{\text{boundary}}$ they will (together) carry a factor of $1/\alpha_\lambda=|G|/d_\lambda$ in each sector $\lambda$. 
Moreover, the action of the dual bases is such as to produce a unit matrix $1_{R_2}$ in the representation space $R_2$ with coefficient equal to the trace in the representation space $R_1$. We need to correct by another factor of $1/d_\lambda$ in order to translate the trace in the space of homomorphisms purely back into a trace over the space including the unit matrix in the representation $\lambda$. We thus find:
\begin{align}
\psi_\alpha \psi \psi^\alpha &= 
%\sum_\mu \frac{|G|}{d_\mu} tr' (P_\mu \psi) 1 \otimes 1
%= 
\sum_\mu \frac{|G|}{d_\mu^2} \text{Tr} (P_\mu \psi) 1_{R_2} \otimes 1_{d_\mu \times d_\mu} \, .
\end{align}
By comparing the two results, we  conclude that $M=1$ for the Cardy condition to hold. The normalization for the bulk trace is therefore $\langle g\rangle_{\text{bulk}} = \frac{1}{|G|} \delta_{1,g}.$\footnote{Consider the following check on this formula. In \cite{Lauda:2006mn}, the bulk trace from the matrix perspective is found to be $d_\mu^2/|G|^2$ for the unit matrix in each $\mu$ sector. When we consider that the identity in the group corresponds to the unit matrix in each $\mu$ sector, we can sum these weights to find $1/|G|$ for the linear form evaluated on unity. } The checks confirm the formulas stated earlier and confirm that we have a consistent open-closed theory \cite{Lazaroiu:2000rk,Moore:2006dw}.

%\begin{equation}
%\langle g\rangle_{\text{bulk}} = \frac{1}{|G|} \delta_{1,g}
%\end{equation}

%This is consistent with equation [LP, 4.33] and the expression for the idempotent, providing factors of $1/|G|$ and $d^2/|G|$. 

\subsubsection{The Group Algebra}
\label{StateSum}
\label{GroupAlgebraPerspective}
Finally, we provide a fourth perspective because we believe it is useful in further clarifying the literature.
For the group theory case, we identified a distinguished theory among all semisimple theories which harbored a particular choice of linear form on the algebra. We will further narrow down our vision by also picking a particular representation $R$ for the open theory, equal to the group algebra $\mathbb{C}[G]$. This choice allows for a very natural construction of the open-closed theory in terms of standard group operations. This fourth perspective is also the occasion to recall some of the underlying mathematics of the state sum reconstructed topological quantum field theories. 

An analysis of these theories was presented in \cite{Lauda:2006mn} in a rather broad framework - we refer once more to \cite{Lauda:2006mn} for background material. Firstly, we restrict their framework to the setting in which their
abelian symmetric monoidal category equals the category $\text{Vect}_{\mathbb{C}}$ of vector spaces over the complex numbers. Secondly, we consider  a strongly separable\footnote{ Strongly separable for us means finite dimensional and with a non-degenerate trace form. For an introduction to Frobenius algebras and topological quantum field theories, see e.g. \cite{Koch}. } symmetric Frobenius algebra object $(A,\mu,\eta,\Delta,\epsilon)$ in the category of vector spaces $\text{Vect}_{\mathbb{C}}$ with (invertible) window element $a$. We recall that $\mu$ represents the product (or joining), $\eta$ the cup (or state-from-nothing), $\Delta$ the co-product (or splitting) and $\epsilon$ the cap (or one-point function).
%(This is automatic ? It is specified in terms of $\mu, \Delta$ and $\eta$ ?)  
There then exists a (particular) knowledgeable Frobenius algebra $(A,C=p(A),i,i^\ast)$ which agrees with the algebra arising from a state space sum definition of an open-closed topological quantum field theory \cite{Lauda:2006mn}.  
We note that the starting point  is an open algebra and it fully determines the closed algebra. 
%
%This reverses perspective compared to the  approaches we presented previously. 

%
As an example of the general construction, we consider the strongly separable symmetric Frobenius algebra $A$ to be the group algebra $A=\mathbb{C}[G]$ \cite{Lauda:2006mn}. The product $\mu$ is simply the (linearly extended) group product
and the state corresponding to an empty disk is the identity $1$ in the group $G$:
\begin{align}
%A &= \mathbb{C} [G]
%\nonumber \\
\mu_A(g \otimes h) &= gh
\nonumber \\
\eta_A(1) &= 1 \, .
\end{align}
We moreover pick the co-product and one-point functions \cite{Lauda:2006mn}:
% \begin{align}
% \eta_{Z(A)} (1) &= \text{wrong in [LP] ?}
% \nonumber \\
% z_\rho &= \frac{d_\rho}{|G|} \sum_{g \in G} \chi_\rho(g) g
% \, .
% \end{align}
% Choice:
\begin{align}
\Delta_A(g) &= \sum_{h \in G} h \otimes h^{-1} g
\nonumber \\
\epsilon_A(g) &= \delta_{g,1} \, .
\end{align}
At this point, the whole construction (including the parameters indicated in previous approaches) is fixed. 
We can  compute the window element \cite{Lauda:2006mn}:
\begin{align}
a=(\mu_A \circ \Delta_A \circ \eta_A) = |G| e = |G| \eta_A(1) \, 
\end{align}
of the algebra $A$ and it exhibits the crucial factor of $|G|$ that entered our boundary trace. The window element defines the idempotent operator $p$ used to project onto the bulk algebra:
\begin{align}
p (g) &= \left( (a^{-1} .\, \text{id}_A) \circ \mu_A \circ \tau_{A,A} \circ \Delta_A \right)(g) \nonumber \\
&= \frac{1}{|G|} \sum_{h \in G} h^{-1} g  h \, ,
\end{align}
where $\tau_{A,A}$ exchanges the two factors of a tensor product.
This projects $g$ onto an element in the center of the algebra which is spanned by sums over conjugacy classes. 
One can compute the bulk Frobenius algebra $(C=p(A),\mu_C,\eta_C,\Delta_C,\epsilon_C)$ from this data. 
One needs the maps:
\begin{align}
\text{im} \, p &: p(A) \rightarrow A : e_\rho \mapsto  \frac{1}{|G|} \sum_{g \in G} d_\rho \chi_\rho(g^{-1}) g
\nonumber \\
\text{coim} \, p &: A \rightarrow p(A) : g \mapsto \frac{1}{|G|} \sum_{h \in G} h^{-1} g h\, .
\end{align}
One finds the bulk multiplication:
\begin{align}
\mu_C (e_{\rho_1} \otimes e_{\rho_2}) 
 &=  \delta_{\rho_1,\rho_2} e_{\rho_1} \, .
\end{align}
We  then determine the co-product $\Delta_C$: 
%We have factors of $|G|$ everywhere. We must then still evaluate $\Delta_C$ using theorem 2.22 as well as $\epsilon_C$ and then compare to the matrix results. We have:
\begin{align}
%\Delta_C &= (\text{coim} \, p \otimes \text{coim} \, p) \circ \Delta_A \circ  a \,  \text{id}_A \circ im \, p 
%\nonumber \\
\Delta_C (e_\rho) &= \left( (\text{coim} \, p \otimes \text{coim} \, p) \circ \Delta_A \circ  a \,  \text{id}_A \right) (e_\rho)
% \nonumber \\
%  &= |G| (coim \, p \otimes coim \, p) \circ \Delta_A \,    (e_\rho)
% \nonumber \\
% &=  (coim \, p \otimes coim \, p) \circ \Delta_A (\sum_{g} d_\rho \chi_\rho(g^{-1})  g)
% \nonumber \\
%  &= (coim \, p \otimes coim \, p ) \sum_{g,h} d_\rho \chi_\rho(g^{-1})  h \otimes h^{-1} g
 \nonumber \\
 &= \frac{1}{|G|^2} \sum_{g,h,h_1,h_2} d_\rho \chi_\rho (g^{-1})
 h_1^{-1} h h_1 \otimes h_2^{-1} h^{-1} g h_2
 \nonumber \\
 % &= \frac{d_\rho}{G} \frac{1}{|G|^2} \sum_{t,g,h,h_1,h_2} d_\rho \chi_\rho ( t^{-1})
 % \chi_\rho (g^{-1} t ) 
 % h_1^{-1} h h_1 \otimes h_2^{-1} h^{-1} g h_2
 % \nonumber \\
 % &= \frac{d_\rho}{G} \frac{1}{|G|^2} \sum_{t,g,h,h_1,h_2} d_\rho \chi_\rho (t^{-1})
 % \chi_\rho (g^{-1} t) 
 % h_1^{-1} t h h_1 \otimes h_2^{-1} h^{-1} t^{-1} g h_2
 % \nonumber \\
 &= \frac{d_\rho^2}{|G|^3} \sum_{t,g,h,h_1,h_2} \chi_\rho (t^{-1})
 \chi_\rho (g^{-1}) 
 h_1^{-1} t h h_1 \otimes h_2^{-1} h^{-1}g h_2
  \nonumber \\
 %  &= \frac{d_\rho}{G} \frac{1}{|G|^2} \sum_{t,g,h,h_1,h_2} d_\rho \chi_\rho (t^{-1})
 % \chi_\rho (g^{-1}) 
 % h_1^{-1} t h h_1 \otimes h_2^{-1} h^{-1}g h_2
&= \frac{1}{|G|} \sum_{h,h_1,h_2} h_{1}^{-1} e_\rho h h_1 \otimes h_2^{-1} h^{-1} e_\rho h_2  
 \nonumber \\
 % &= \frac{d_\rho}{G} \frac{1}{|G|^2} \sum_{t,g,h,h_1,h_2} d_\rho \chi_\rho (t^{-1})
 % \chi_\rho (g^{-1} ) 
 % h_1^{-1} h h_1 \otimes h_2^{-1} h^{-1} tg h_2
 % \nonumber \\
 % &= \frac{d_\rho}{G} \frac{1}{|G|^2} \sum_{t,g,h,h_1,h_2} d_\rho \chi_\rho (h^{-1} t^{-1})
 % \chi_\rho (g^{-1} ) 
 % h_1^{-1} h h_1 \otimes h_2^{-1} tg h_2
 % \nonumber \\
 % &= \frac{1}{|G|^2} \sum_{g,h,h_1,h_2} d_\rho \chi_\rho (g^{',-1} h^{-1} )
 % h_1^{-1} h h_1 \otimes h_2^{-1} g' h_2
 % \nonumber \\
 &= \frac{|G|^2}{d_\rho^2} e_\rho \otimes e_\rho
\end{align}
We exploited the reverse of the identity to prove that $e_\rho^2=e_\rho$:
\begin{equation}
\sum_{t \in G} \chi_\rho(t^{-1} ) \chi_\rho(g^{-1} t)
=  \frac{|G|}{d_\rho} \chi_\rho(g^{-1}) \, .
\end{equation}
In the last lines, we use that the $e_\rho$ are orthogonal and central idempotents to show that each tensor factor is proportional to $e_\rho$. The coefficient can be computed by taking traces. One needs $\text{tr}( e_\rho h )= \chi_\rho(h)/d_\rho \text{tr} (e_\rho) $.  Applying this twice and using the orthogonality relation for characters once more, one obtains the end result.
We  also compute the one-point function:
\begin{align}
%\epsilon_C &= \epsilon_A \circ a^{-1} \text{id}_A \circ im \, p
%\nonumber \\
\epsilon_C (e_\rho) &= (\epsilon_A \circ a^{-1} \text{id}_A) (\frac{1}{|G|} \sum_g d_\rho \chi_\rho (g^{-1}) g)
\nonumber \\
&= \frac{1}{|G|^2} d_\rho \chi_\rho (1) = \frac{d_\rho^2}{|G|^2} \, .
\end{align}
These results are precisely as in the group theoretic or matrix algebra perspective provided we consider the representation space $R=\mathbb{C}[G]$ which is equivalent to $R=d_\lambda \lambda$ where we sum over irreducible representations $\lambda$ of $G$ with multiplicity equal to their dimension. 
%This theory reproduces all of the finite group topological quantum field theory if we assume we have a "regular" D-brane.
%
Since we have a state sum reconstructed theory, we can  compute the weight $g^{(3)}$ corresponding to a triangular plaquette. It is equal to \cite{Lauda:2006mn}
\begin{align}
%g^\ast(1) &= \sum_h h \otimes h^{-1}
%\nonumber \\
g^{(3)} ((g \otimes h) \otimes l) &=(\epsilon_A \circ \mu_A (\mu_A \otimes \text{id}_A))((g \otimes h) \otimes l)= 1 \quad \text{iff} \quad ghl=1 \, ,
\end{align}
namely a plaquette contributes if the surrounding 
group links multiply to one. 
Specifying the full lattice theory with boundary is algorithmic given the data of the Frobenius algebra $A$ \cite{Lauda:2006mn}.

\subsubsection*{Remarks}
We offer some final remarks on the comparison of our review with the literature and of the individual resources in the literature with each other. The treatment of the open-closed theory associated to a finite group in \cite{Moore:2006dw,Moore} is complete for one choice of boundary or bulk trace. The theory in \cite{Lauda:2006mn} allows for more general bulk metric choices in the matrix algebras. On the other hand, while treating the group example, \cite{Lauda:2006mn} restricts  the setting of \cite{Moore:2006dw,Moore} to the case of the representation $R=\mathbb{C}[G]$. We conclude that we can allow for a generalization of \cite{Lauda:2006mn,Moore:2006dw,Moore} in which we pick more general  conjugation invariant bulk one-point functions and allow for generic group representations. These define open-closed theories that fit into the matrix framework of \cite{Lauda:2006mn}. Compared to \cite{Lauda:2006mn}, we  clarified that the choice of window element  can be thought of as allowing for a boundary specified by the multiplicities of the irreducible representations (given a choice of parameters for the bulk metric). The choice of bulk one-point functions in \cite{Moore:2006dw,Moore} is canonical for various reasons\footnote{For instance,  the delta-function on the group  is the unit for convolution.} but the other (finite group) theories (with boundary) are also consistent when provided with more general conjugation invariant measures.\footnote{Namely, a more general choice of trace parameters $\alpha_\lambda$.} All these theories can be state sum reconstructed \cite{Lauda:2006mn}. The semisimple perspective is  powerful but may obscure some of the underlying group structure that one often prefers to keep manifest. Vice versa, too much of a group perspective can cloak the possibilities that the semisimple matrix perspective opens up.

\section{The Finite Inverse Monoid  Theory}
\label{FiniteInverseMonoids}
\label{MonoidTQFT}
In this section, we associate an open-closed topological quantum field theory to a finite inverse monoid. A monoid $M$ is a set equipped with an associative multiplication with an identity element. It is finite if the cardinal number of the set is finite. The monoid is inverse if for every element $m \in M$ there is an element $m^\ast$ such that $m \, m^\ast \, m=m$ and $m^\ast \, m m^\ast =m^\ast$. As such, it is a generalization of a finite group $G$ (in which the star is the inverse).  We construct an associated open-closed topological quantum field theory in analogy to the group case discussed in section \ref{GroupTQFT}. To see the analogy, we need two crucial facts on the structure theory of finite inverse monoids.  Firstly, to each finite inverse monoid there is a canonically associated groupoid \cite{Lawson,Steinberg} and their algebras are isomorphic. Secondly, the groupoid algebra is equivalent to a matrix algebra. Thus, the finite inverse monoid algebra $\mathbb{C} [M]$ is semisimple. We recall these rather abstract facts here in a manner that is hopefully both efficient and transparent. Once we have established the isomorphisms, the state sum reconstructed open-closed theories for matrix algebras can be exploited to formulate the monoid theories. 

The section is planned  as follows. In subsection \ref{StructureTheory} we recall the structure theory of finite inverse monoids, their representation theory and how to associate a groupoid to each monoid \cite{Steinberg}.\footnote{All monoids are assumed to be finite and inverse in this paper. Some properties are more generally valid. We refer to \cite{Steinberg} for a pedagogical introduction with references to the original literature. } 
%In subsection \ref{GroupoidTQFT} we map a groupoid to a topological quantum field theory. 
Subsection \ref{SubsectionMonoidTQFT}  contains the construction of the open-closed topological quantum field theory for monoids.

\subsection{The Structure Theory of Finite Inverse Monoids}
\label{StructureTheory}
We review properties of (finite inverse) monoids, the monoid algebra and the associated groupoid algebra. We restrict to the characteristics that will be of use in the construction of an associated open-closed topological quantum field theory. 
Our main reference throughout is the textbook \cite{Steinberg}
which we recommend for both  proofs and details.

\subsubsection{Monoids,  Partially Ordered Sets and Maximal  Subgroups}
Firstly, we wish to recall that,  loosely speaking, the structure of a monoid $M$ is that of a set of finite groups glued together by a partially ordered set of idempotents \cite{Steinberg}.
%An element of a monoid that has an inverse is called a unit. The units form a group that we denote $G_1$. 
An element $e$ of the monoid is idempotent if $e^2=e$. The set of idempotents of the monoid $M$ is denoted $E(M)$. 
%Note that the identity of $eMe$ is $e$. 
There is a natural partial order on the set of idempotents which is $e \le f$ if $ef=e=fe$. To each idempotent $e$, we associate the group $G_e$ which is the set of invertible elements of $eMe$. 
In an inverse monoid we have that the groups $G_e$ can alternatively be described as
\begin{equation}
G_e = \{ m \in M | m^\ast m = e = m m^\ast \}
\, .
\end{equation}
There is in fact a natural partial order on the full inverse monoid. The partial order is
$m \le n$ if $m=nm^\ast m$.
There is also an equivalence relation ${\cal J}$  among elements of monoids which is 
\begin{equation}
m_1 \, {\cal J} \, m_2 \quad \text{if and only if} \quad Mm_1M=Mm_2M \, .
\end{equation}
This divides up the monoid $M$ into ${\cal J}$ equivalence classes. 
%It is regular when all its elements satisfy $m \in  m M m$. 

\subsubsection{The Representations
%, the Class Functions 
and the Algebra}
One can build the representations of the monoid $M$ from the representations of the groups $G_e$
associated to the idempotents $e \in E(M)$. 
In particular, there is a one-to-one map from the irreducible representations of a (finite inverse) monoid to the irreducible representations of its  subgroups $G_{e_i}$, where one takes one idempotent element $e_i$ per 
%regular 
${\cal J}$ equivalence class. 

Importantly, the monoid algebra $\mathbb{C}[M]$ for a finite inverse monoid is semisimple. This has as a consequence that its modules are  completely reducible. 
Since the algebra $\mathbb{C}[M]$ is semisimple, it is isomorphic to a matrix algebra over the complex numbers. If there are $n_i$ idempotent elements in the ${\cal J}$ class of $e_i$, then the algebra is isomorphic to the matrix algebra:
\begin{equation}
\mathbb{C}[M] \cong \sum_{e_i} M_{n_i} (\mathbb{C} [G_e]) \label{MonoidAlgebraMatrixAlgebra}
\end{equation}
where the sum is over idempotent representatives $e_i$ of ${\cal J}$ equivalence classes. 

\subsubsection{The Groupoid}
It is useful to introduce an abstract object which is a groupoid. 
The groupoid algebra will be a stepping stone that allows us to describe the isomorphism (\ref{MonoidAlgebraMatrixAlgebra}) between the monoid algebra and the matrix algebra explicitly. This is instrumental in describing the topological field theory hands on. Let us therefore momentarily dive into the category theory for groupoids.

The groupoid $\text{Go}(M)$ associated to a finite inverse monoid $M$ is a category with objects in the set of idempotents $E(M)$ and an arrow $[m]$ associated to each element $m$ of the monoid. 
%(We just use a different notation for it ..) 
One has  the source map $s([m])=m m^\ast$ and the target map $t([m])=m^\ast m$ which associates two objects to each arrow. The subtle point is that the composition of arrows is different from the monoid product. We have that $[m][n]=[mn]$ if and only if $t([m])=s([n])$. Otherwise, the product is zero. For the category, this makes sense, as arrows only compose when they are consecutive.  The identity associated to the idempotent object $e$ is $1_e=[e]$ and the inverse arrow equals $[m]^{-1}=[m^\ast]$. We understand that the groupoid structure is closer to a group than is the monoid. Hence, the intermediate nature of the groupoid stepping stone.  We do  keep in mind that the product is different in groupoid and monoid. 

There are two important isomorphisms that we wish to exhibit. Firstly, the monoid and the groupoid algebras are isomorphic. Secondly, the groupoid algebra is isomorphic to a (direct sum of) matrix algebra(s). After these two steps, we will have an explicit description of the isomorphism (\ref{MonoidAlgebraMatrixAlgebra}). 
The first  isomorphism $\alpha$ is between the algebra $\mathbb{C}[M]$ of the  monoid and the algebra $\mathbb{C}[\text{Go}(M)]$ of the groupoid:
\begin{align}
\alpha: \mathbb{C}[M] \rightarrow \mathbb{C}[\text{Go}(M)]: m \mapsto \alpha(m) &= \sum_{ n \le m} [n]
\end{align}
with inverse $\beta=\alpha^{-1}$:
\begin{align}
\beta: \mathbb{C}[\text{Go}(M)] \rightarrow \mathbb{C}[M]  : [m]  \mapsto  
\beta([m])= \sum_{n \le m} \mu(n,m) n \, .
\end{align}
It is given in terms of the partial order $(M,\le)$ on the monoid and its corresponding 
%(Theorem 9.3) 
 M\"obius function $\mu$.\footnote{The isomorphism is a consequence of the following property of the product of elements in inverse monoids:  each monoid element $n \le m_1 m_2$ is uniquely factorized as $n=n_1 n_2$ where $n_i \le m_i$ and $n_1^\ast n_1 = n_2 n_2^\ast$. The full proof is in \cite{Steinberg}. } 
Secondly, the groupoid algebra $\mathbb{C}[\text{Go}(M)]$ is semisimple and 
equivalent to the matrix algebra
\begin{equation}
\mathbb{C} [\text{Go}(M)] \cong \prod_{i=1}^s M_{n_i} (\mathbb{C} [G_{e_i}])
\, ,
\end{equation}
where $e_1,\dots,e_s$ are representatives of the isomorphism classes of objects in the groupoid, namely  the idempotent representatives of ${\cal J}$ equivalence classes in the monoid. To understand the second isomorphism better, we first concentrate on one factor. 
In each factor all objects $x_m$ (among which is $e_i$) are connected.  Choose then arrows $p_m: x_m \rightarrow x_1$ which link up  object $x_m$ to a chosen object $x_1$ (e.g. equal to $e_i$). Define the isomorphism $\alpha': \mathbb{C} [\text{Go}(M)] \rightarrow M_n (\mathbb{C} G_{x_1})$ on an arrow $[n]: x_m \rightarrow x_n$ by:
\begin{equation}
\alpha'([n]) = (p_m^{-1} [n] \, p_l) e_{ml} \, .
\end{equation}
Note that the coefficient is indeed an arrow from $x_1$ to itself that is therefore a group element. 
The inverse isomorphism for a summand is:
\begin{equation}
\beta'((a_{lm})) = \sum_{l,m} p_l a_{lm} p_m^{-1} \, .
\end{equation}
It extends to the full algebra through a decomposition of unity \cite{Steinberg}. 
The composition of the isomorphisms $\alpha$ and $\alpha'$ provides the explicit relation between the monoid algebra $\mathbb{C}[M]$ and the manifestly semisimple matrix group algebra:
\begin{equation}
\mathbb{C} [M] \stackrel{ \alpha}{\cong} 
\mathbb{C}[\text{Go}(M)]
\stackrel{\alpha'}{\cong}
\prod_{i=1}^s M_{n_i} (\mathbb{C} G_{e_i})
\, .
\label{isomorphism}
\end{equation}

\subsubsection{The Irreducible Representations}
The isomorphism (\ref{isomorphism}) demonstrates that the 
 representation theory of the inverse monoid $M$ (or equivalently, its algebra $\mathbb{C}[M]$) is indeed inherited from the representation theory of the maximal groups $G_{e_i}$. Concretely, an irreducible representation of $G_{e_i}$ can be tensored with the (unique irreducible) $n_i$ dimensional representation of  $M_{n_i}(\mathbb{C})$ to become a representation of the matrix algebra $M_{n_i}(\mathbb{C} G_{e_i})$. We obtain a full irreducible representation of the product of matrix algebras by taking the representation to be zero for the other factors. The result is a bijection (through the isomorphism) of the irreducible representations of the monoid  to the irreducible representations of the groups $G_{e_i}$. The dimension of the irreducible representation equals $n_i d$ where $d$ is the dimension of the irreducible representation of the group $G_{e_i}$. The sum of the dimensions squared of the irreducible representations adds up to the dimension of the monoid.

\subsubsection{The Central Idempotents}
In the matrix algebra description of the monoid algebra, it is clear that the central idempotents correspond to an identity matrix in a factor matrix algebra multiplied by an idempotent of the group algebra. Explicitly, consider a central idempotent of the group $G_e$ associated to a representative $e$ of a ${\cal J}$ class and an irreducible representation $\rho$ of $G_e$. Inside the algebra $\mathbb{C}[G_e]$, we write the idempotent:
\begin{equation}
e_{e,\rho} = \frac{d_{e,\rho}}{|G_e|} \sum_{g \in G_e} \chi_\rho (g^{-1}) g \, .
\end{equation}
It is inherited by the block $M_{n}(\mathbb{C} [G_{e}])$ inside the matrix algebra by multiplying with the unit matrix in the  idempotent block associated to the $\mathcal{J}$ equivalence class $J$.
The central idempotent of the matrix algebra is then:
\begin{equation}
e_{e,\rho} = \frac{d_{e,\rho}}{|G_e|} \sum_{g \in G_e} \chi_\rho (g^{-1}) g \otimes 1_{n_e \times n_e} \, . 
\end{equation}
To obtain the central idempotents in the monoid algebra, we have to pull this result through the inverse isomorphisms $\beta'$ and $\beta$ -- we abuse notation on the left-hand side --:
\begin{align}
e_{e,\rho} &= ( \beta \circ \beta') \left(\frac{d_{e,\rho}}{|G_e|} \sum_{g \in G_e} \chi_\rho (g^{-1}) g \otimes 1_{n_e \times n_e} \right)
\nonumber \\
&= \beta \left( \sum_{f \in E(J)} \frac{d_{e,\rho}}{|G_e|} 
\sum_{g \in G_e} \chi_\rho (g^{-1}) p_f [g]  p_f^{-1} \right)
\nonumber \\
&= \sum_{f \in E(J)} \frac{d_{e,\rho}}{|G_e|} 
\sum_{g \in G_e} \chi_\rho (g^{-1})
\sum_{n \le p_f g  p_f^{-1} }   \mu(n,p_f g  p_f^{-1}) \, n
\, .  \label{MonoidIdempotent}
\end{align}
In the second line, it is understood that $p_f$ is an arrow from idempotents $f$ to $e$ and $g$ plays  the role of a group element associated to the idempotent $e$. It is a consequence of the fact that the unit matrix is the sum of identity maps associated to each idempotent $f$ in $E(J)$. The maps $p_f$ implement the isomorphism between the groups $G_e$ and $G_f$. 
In the last line $p_f g  p_f^{-1}$ is the monoid element that corresponds to the arrow from $f$ to $f$ associated to the group element $g$.\footnote{
These idempotents were first written down  in \cite{Steinberg06}. See also appendix \ref{IdempotencyAppendix} for an operational proof of idempotency.}

\subsection{The Open-Closed Monoid Theory}
\label{SubsectionMonoidTQFT}
The open-closed monoid topological quantum field theory can again be viewed from multiple perspectives. Firstly, the monoid algebra is semisimple. From its equivalence to a matrix algebra, we conclude that its center has dimension equal to the sum over ${\cal J}$ equivalence classes labeled by $e_i$ of the number of conjugacy classes of the groups $G_{e_i}$. Therefore, we have that the center $Z(\mathbb{C}[M]) \cong \mathbb{C}^{\sum r_i}$ where $r_i$ is the number of conjugacy classes of $G_{e_i}$. The semisimple abstract view on the theory is then as stated in subsection \ref{SemisimplePerspective} with the number of points of $X=\text{Spec}(Z(\mathbb{C}[M]))$ equal to $\sum r_i$. As in section \ref{GroupTQFT}, we explore different perspectives on the open-closed monoid theories and identify a choice of theory inspired by an algebraic set-up.

\subsubsection{The Groupoid Algebra Perspective}
\label{GroupoidAlgebraPerspective}
\label{GroupoidTQFT}
We presented a group algebra open-closed theory in subsection \ref{GroupAlgebraPerspective}. In this subsection, we take the stepping stone groupoid algebra in the equivalence relations (\ref{isomorphism}) and build an analogous theory that comes with particular choices for the linear form and the spectrum of boundary operators. We build it such that it reduces to the group case when the groupoid is a group.\footnote{The theory we study is different from example 5.1 in \cite{Lauda:2006mn} in the choice of linear form. As for the group, we propose to introduce a non-trivial window element namely a non-canonical Frobenius algebra in the nomenclature of \cite{Lauda:2006mn}.}
%
% \subsubsection*{Groupoids}
% Intermezzo: we can make remarks on general monoids here (or elsewhere). They are non-trivial and may be of interest.
%
The groupoid algebra $A=\mathbb{C} [\text{Go}(M)]$ is again an example of a strongly separable symmetric Frobenius algebra.
We  have a large choice for the linear form. Our choice naturally flows from the following definitions. 
We drop the square brackets on the groupoid arrows for ease of notation and define:
\begin{align}
\Delta_A(g) &= \sum_{h \in \text{Go} | s(h)=s(g)} h \otimes (h^{-1} g)
\nonumber \\
\epsilon_A(g) &= 1 \quad \text{iff} \quad g = 1_{s(g)} \, . 
\end{align}
The last line says that the groupoid arrow $g$ must be the identity at its source in order to have a non-zero one-point function. 
That implies it also has equal source and target.
% (This indeed corresponds to an element on the diagonal in the matrix picture and the normalization seems to be coherent as well.)
Given these  formulas, we can be more explicit about our chosen state sum theory for (finite inverse) monoids, or rather its associated groupoid algebra. 
We first compute the window element: 
\begin{equation}
a = (\mu_A \circ \Delta_A \circ \eta_A)(1) = (\mu_A \circ \Delta_A) \sum_{ e \in E(M)} 1_e= \sum_{e,h \in \text{Go}| s(h)=e} 1_e = \sum_{e \in E(M)} n_e  |G_e| 1_e \, .
\end{equation}
We find a factor of $n_e |G_e|$ for each idempotent because that is the number of arrows with source equal to a given idempotent $e$.  The factor depends on the ${\cal J}$ equivalence class of the idempotent only. 
%(If we count per all idempotent, there is not factor of $n_e$).)
The construction of the knowledgeable algebra (namely, the open-closed theory) uses the  idempotent projection:
\begin{align}
p(m) &= (a^{-1} \text{id}_A \circ \mu_A \circ \tau_{A,A} \circ \Delta_A) (m)= \delta_{s(m),e} \frac{1}{ n_e|G_e|} \sum_{t(m)=s(h)=s(m)=e} h^{-1} m h \, 
%\, \text{if} \, \, t(m)=s(m) \, 
. \nonumber 
\end{align}
In other words, for each arrow from an object to itself, we project the arrow onto its groupoid invariant counterpart. Other arrows are projected to zero. We conjugate both to obtain a group invariant and to spread the groupoid arrow over the ${\cal J}$ block along the diagonal in the idempotent space $E(J_e)$. The normalization of the projection operator is therefore appropriate. 
We calculate the ensuing product $\mu_C$, co-product $\Delta_C$ and one-point function $\epsilon_C$ for the bulk Frobenius algebra that results from the projection using the formalism of \cite{Lauda:2006mn}. We  need the maps:
% Differs from [Lauda]. 
\begin{align}
\text{im} \, p &: p(A) \rightarrow A : 
 e_{e,\rho}  \mapsto \frac{d_{e,\rho}}{|G_e|} \sum_{g \in G_e} \chi_\rho (g^{-1}) g \otimes 1_{n_e \times n_e} 
\\
\text{coim} \, p &: A  \rightarrow p(A):  g \mapsto
\frac{1}{n_{s(g)} G_{s(g)}}
\sum_h h^{-1} g  h
\quad {\text{for}} \quad s(g)=t(g) \quad  \text{and}  \quad s(g)=s(h) \, . \nonumber 
\end{align}
We have that $e_{e,\rho}$ are idempotents  and therefore the closed product $\mu_C$  is:
\begin{align}
\mu_C (e_{e,\rho} \otimes e_{e,\mu}) &= \delta_{\rho,\mu} e_{e,\rho} \, .
\end{align}
 Next, we compute the linear form $\epsilon_C$:
\begin{align}
\epsilon_C (e_{e,\lambda}) &= (\epsilon_A \circ a^{-1} \text{id}_A \circ \text{im} \, p)(e_{e,\lambda})
\nonumber \\
%\epsilon_C (e_{e,\lambda}) 
% &= (\epsilon_A \circ a^{-1} \text{id}_A) \left( \frac{d_{e,\lambda}}{|G_e|} \sum_{g \in G_e} \chi_\lambda(g^{-1}) g \otimes 1_{n_e \times n_e} \right)
% \nonumber \\
&= \epsilon_A  \left( \frac{d_{e,\lambda}}{n_e |G_e|^2} \sum_{g \in G_e} \chi_\lambda(g^{-1}) g \otimes 1_{n_e \times n_e} \right)
\nonumber \\
&= \frac{d_{e,\lambda} \chi_\lambda (e)}{|G_e|^2}  = \frac{( d_{e,\lambda})^2}{|G_e|^2} \, .
\end{align}
We also calculate the co-product $\Delta_C$:
\begin{align}
% \Delta_C &= (coim \, p \otimes coim \, p) \circ \Delta_A \circ  a \,  \text{id}_A \circ im \, p 
% \nonumber \\
\Delta_C (e_{e,\rho}) &= \left((\text{coim} \, p \otimes \text{coim} \, p) \circ \Delta_A \circ  a \,  \text{id}_A \right) (e_{e,\rho})
% \nonumber \\
%  &=n_e |G_e| \left( (\text{coim} \, p \otimes \text{coim} \, p) \circ \Delta_A \right) \,    (e_{e,\rho})
\nonumber \\
&= n_e \left(  (\text{coim} \, p \otimes \text{coim} \, p) \circ \Delta_A  \right)
(\sum_{s \in G_e} d_{e,\rho} \chi_\rho(g^{-1})  g 1_{n_e \times n_e})
\nonumber \\
 &= n_e (\text{coim} \, p \otimes \text{coim} \, p ) \left(\sum_{f \in E(J_e)} \sum_{g \in G_e, h|s(h)=f} d_{e,\rho} \chi_\rho(g^{-1})  h \otimes h^{-1} (g 1_f)\right)
\, .
 %t^{-1}  \mu(t,s)
 % \nonumber \\
 % &= \frac{1}{n_e |G_e|^2} \sum_{f \in E(J_e), g\in G_e,h|s(h)=f,h_1|s(h_1)=s(h)=f,h_2|s(h_2)=t(h)=t(g)=f} d_{e,\rho} \chi_\rho (g^{-1})
 % h_1^{-1} h h_1 \otimes h_2^{-1} h^{-1} g 1_f h_2 \nonumber 
 \end{align}
 Our initial element $g$ corresponds to  group elements combined with $1_f$ for the whole ${\cal J}$ equivalence class of idempotents.  That is the origin of the sum over the idempotents $f$ in the equivalence class $J_e$ of the representative idempotent $e$. When we apply the co-image in the following line, we separately do this for the first and last factor.  This will imply that $h$ also must be diagonal because of the first factor and it must agree with $g 1_f$ in idempotent space because of the last factor. We can then compute and perform changes of variables on $g,h$ in terms of the group  $G_f \cong G_e$:
 \begin{align}
 \Delta_C (e_{e,\rho}) &=
  \frac{d_{e,\rho}^2}{n_e |G_e|^3}  \sum_{f,k,g,h,h_1,h_2}  \chi_\rho ( k^{-1})
 \chi_\rho (g^{-1} k ) 
 h_1^{-1} h h_1 \otimes h_2^{-1} h^{-1} g  1_f
 %\mu(t,s) 
 h_2
 % \nonumber \\
 % &= \frac{d_{e,\rho}}{G_e} \frac{1}{n_e |G_e|^2} \sum_{k,g,h,h_1,h_2} d_{e,\rho} \chi_\rho (k^{-1})
 % \chi_\rho (g^{-1} k) 
 % h_1^{-1} k h h_1 \otimes h_2^{-1} h^{-1} k^{-1} g  1_f %\mu(t,s) t^{-1}
 % h_2
 \nonumber \\
 &=  \frac{d_{e,\rho}^2}{n_e |G_e|^3}  \sum_{f,g,k,h,h_1,h_2}  \chi_\rho (k^{-1})
 \chi_\rho (g^{-1}) 
 h_1^{-1} 
 %\mu(\dots,\dots) 
 k h h_1 \otimes h_2^{-1} h^{-1} g 1_f %\mu(t,s) t^{-1}
 h_2
  \nonumber \\
 %  &= \frac{d_{e,\rho}}{G} \frac{1}{|G|^2} \sum_{t,g,h,h_1,h_2} d_{e,\rho} \chi_\rho (t^{-1})
 % \chi_\rho (g^{-1}) 
 % h_1^{-1} t h h_1 \otimes h_2^{-1} h^{-1}g h_2
&= \frac{1}{n_e^2 |G_e|} \sum_{h,h_1,h_2} h_{1}^{-1} e_{e,\rho} h h_1 \otimes h_2^{-1} h^{-1} e_{e,\rho} h_2  
 \nonumber \\
 % &= \frac{d_{e,\rho}}{G} \frac{1}{|G|^2} \sum_{t,g,h,h_1,h_2} d_{e,\rho} \chi_\rho (t^{-1})
 % \chi_\rho (g^{-1} ) 
 % h_1^{-1} h h_1 \otimes h_2^{-1} h^{-1} tg h_2
 % \nonumber \\
 % &= \frac{d_{e,\rho}}{G} \frac{1}{|G|^2} \sum_{t,g,h,h_1,h_2} d_{e,\rho} \chi_\rho (h^{-1} t^{-1})
 % \chi_\rho (g^{-1} ) 
 % h_1^{-1} h h_1 \otimes h_2^{-1} tg h_2
 % \nonumber \\
 % &= \frac{1}{|G|^2} \sum_{g,h,h_1,h_2} d_{e,\rho} \chi_\rho (g^{',-1} h^{-1} )
 % h_1^{-1} h h_1 \otimes h_2^{-1} g' h_2
 % \nonumber \\
 &= \frac{|G_e|^2}{d_{e,\rho}^2} e_{e,\rho} \otimes e_{e,\rho} \, .
\end{align}
The third line uses on the one hand the sum over the idempotent $f$ to restore the unit matrix in the idempotent $e_{e,\rho}$ in one factor, and in a second factor we must divide by $n_e$ in order to obtain the unit matrix in the idempotent space. In the last line, 
we  used the identity
\begin{equation}
tr(e_{e,\rho} h) = tr(e_\rho) \chi_{\rho}(h)/d_{e,\rho} \, ,
\end{equation}
twice
and the sums over $h_1$ and $h_2$  contribute factors of $n_e |G_e|$ while the sum over $h$ is responsible for a factor of $|G_e|$. For illustration purposes, we again compute the plaquette weight in the lattice theory:
\begin{align}
% (\mu \circ \Delta \circ \eta)(1) &= (\mu \circ \Delta) \sum_e 1_e
% \nonumber \\
% &= \mu ( \sum_e \frac{1}{N_{t(1_e)}} \sum_{h: s(h)=s(e)=e} h \otimes h^{-1}) (?)
% \nonumber \\
% &= \sum_e \frac{1}{N_{t(1_e)}} \sum_{h: s(h)=s(e)=e} 1_e
% \nonumber \\
% &= \sum_e \frac{1}{N_{t(1_e)}}  N_{s(e)} 1_e
% \nonumber \\
% &= \sum_e 1_e (? \text{clearly wrong; cfr. group})
% \nonumber \\
% & = ?? \sum_ e |G_e| 1_e ??
% \nonumber \\
% g^\ast(1) &= (\Delta \circ \mu)(1) 
% \nonumber \\
%  &=  \sum_e \sum_{h \in G_e} h \otimes h^{-1} 
% \nonumber \\
g^{(3)} ((g \otimes h) \otimes l)  &=(\epsilon_A \circ \mu_A (\mu_A \otimes \text{id}_A))((g \otimes h) \otimes l)
\nonumber \\
 &= 1 \quad \text{iff } \quad ghl= 1_e  \quad \text{in the groupoid}
\, .
\end{align}
% The invariants of the semisimple algebra correspond to the one-point functions of the idempotents: 
% \begin{align}
% \langle e_{e,\rho} \rangle_{\text{bulk}} &=  \frac{(d_{e,\rho})^2}{|G_e|^2}  \, .
% \end{align}
We performed this analysis on the groupoid stepping stone. It is tedious but straightforward to apply the inverse isomorphism $\beta=\alpha^{-1}$ to formulate the theory directly in terms of the monoid elements. The M\"obius function will decorate our trove of formulas. 
\subsubsection{The Monoid Representation Theory and Matrix Algebra Perspective}
The representation theoretic and the matrix algebra view on the open-closed theory are analogous to the group case once more, but each representation of the group $G_e$ appears $n_e$ times. Thus, we work in terms of the algebra of homomorphisms of irreducible representations with a factor of a big unit matrix attached:
\begin{equation}
A = \sum_{e,\mu} M_{m_{e,\mu}} (\mathbb{C}) \otimes \{ 1_{d_{e,\mu} \times d_{e,\mu} } \otimes 1_{n_e \times n_e} \} 
\, .
\end{equation}
Namely, the dimension of the irreducible representation $R_{(e,\mu)}$ equals $n_e d_{e,\mu}$. 
We have an extra identity because the irreducible representations of the algebra
\begin{equation}
M_{n_e} (\mathbb{C} [G_e]) \cong M_{n_e} (\mathbb{C}) \otimes \mathbb{C}[G_e]
\end{equation}
have the $n_e$ dimensional irreducible representation of $M_{n_e} (\mathbb{C})$ tensored in. 
We again have the central idempotents:
\begin{equation}
z_{e,\mu} = \sum_{p=1}^{m_{e,\lambda}} e_{pp}^{(e,\lambda)}
\end{equation}
tensored with identities and
 the boundary trace has parameters $\alpha_{e,\mu}$ corresponding to traces in the summands in the algebra:
\begin{equation}
\langle e_{pq}^{(e,\lambda)} \rangle_{\text{boundary}} = \delta_{p,q} \alpha_{e,\lambda} \, .
\end{equation}
The window element equals:
\begin{equation}
a = \sum_{e,\mu} \frac{m_{e,\mu}}{\alpha_{\mu,e}} z_{e,\mu} \,.
\end{equation}
We have to make a choice for the parameters $\alpha_{e,\mu}$. We fix:
\begin{equation}
\alpha_{e,\mu} = \frac{d_{e,\mu} %n_e
}{|G_e|} \, .
\end{equation}
The parameter choice can be thought of as a factor of $n_e d_{e,\mu}$ for the dimension of an irreducible representation space and a denominator of $n_e |G_e|$ for the number of monoid elements involved in averaging.
%The bulk one-point function  evaluated on the central idempotents $e_{e,\mu}$ equals $\alpha^2_{e,\mu}$.   
We have the usual closed-open embedding map $i$:
\begin{equation}
i( \sum_{e,m} c^{e,\mu} z_{e,\mu}) =  \sum_{e,m} c^{e,\mu}
\sum_{p=1}^{m_{e,\lambda}} e_{pp}^{(e,\mu)} \, 
\end{equation}
and given the linear forms, its adjoint $i^\ast$ is fixed. 
We compute the relevant coefficients.  On the boundary side, we find:
\begin{equation}
\langle i(c^{e,\mu} z_{e,\mu}) \psi) \rangle_{\text{boundary}}
=c^{e,\mu} \frac{1}{n_e |G_e|} 
%m_{e,\mu} 
\text{Tr} (P_{e,\mu} \psi)  = c^{e,\mu} \alpha_{e,\mu} \text{tr'}
(P_{e,\mu} \psi)
\end{equation}
where the big trace contains factors of $d_{e,\mu}$ and $n_e$ while the primed trace is only over the matrix space of homomorphisms.
The bulk counterpart reads:
\begin{align}
\langle 
i^\ast(\psi) c^{e,\mu} z_{e,\mu} \rangle_{\text{bulk}} &=  \sum_{f,\lambda}  M_{f,\lambda} \text{tr}' (P_{f,\lambda} \psi) c^{e,\mu}
\langle z_{f,\nu}  z_{e,\mu} \rangle_{\text{bulk}}
\nonumber \\
&=  \sum_{e,\mu} M_{e,\mu} c^{e,\mu} \text{tr}' (P_{e,\mu} \psi) \alpha_{e,\mu}^2
\end{align}
We summarize that:
\begin{align}
% i (c^{e,\mu} z_{e,\mu}) &=  c^{e,\mu} \sum_{p=1}^{m_{e,\lambda}} e_{pp}^{(e,\mu)}
% \nonumber \\
i^\ast(\psi) &= M_{f,\nu} \, \text{tr}'(P_{f,\nu} \psi) z_{f,\nu}
\end{align}
with coefficient:
\begin{align}
M_{f,\nu} &= \alpha^{-1}_{f,\nu} \, . \label{MParameter}
\end{align}
As in the group theory case, when this condition is met, the Cardy constraint is also satisfied. Of course, the multiplicities $m_{e,\lambda}$ of the irreducible representation spaces is general in this approach (which is therefore more general than the description in subsection \ref{GroupoidAlgebraPerspective}). 

\subsubsection*{Entries in the  Dictionary}
The formulation of the open-closed finite inverse monoid theory is  complete. We presented the equations mostly in a matrix algebra or groupoid language. The isomorphisms (\ref{isomorphism}) tells us that there is in principle one more step to take to reformulate all equations directly in terms of the inverse monoid. We execute this tedious exercise only to the degree that it may clarify the theory further.  

We have the central idempotent\footnote{See equation (\ref{MonoidIdempotent}). We took some liberty with the notation to write the formula more compactly. See also \cite{Steinberg06}.}
\begin{equation}
e_{e,\rho} = \sum_{e \in E(J)} \left( \frac{\chi_\rho(e)}{|G_e|} \sum_{s \in G_e} \chi_\rho(s^{-1}) \sum_{n \le s} n \mu (n,s) \right) \, ,
\end{equation}
that can be used to decompose a generic representation.  There are as many of these idempotents as there are $J$ classes and irreducible representations of the maximal groups associated to those classes.  Since these idempotents are central, they are in the bulk algebra.
 The image of a central idempotent in a representation is the representation of this element. Moreover, idempotency and centrality is preserved because the map $i$ is a homomorphism.  Thus, an idempotent $e_{e,\mu}$ is mapped into the unit matrix associated to $(e,\mu)$.
We therefore have:
\begin{equation}
i(e_{e,\mu}) =  \sum_{e \in E(J)} \left( \frac{\chi_\mu (e)}{|G_e|} \sum_{S \in G_e} \chi_\mu(s^{-1}) \sum_{n \le s} R(n) \mu (n,s) 
\right) 
\cong \sum_p e_{pp}^{(e,\mu)} \, .
\end{equation}
We  note that the one-point function on the monoid is quite intricate. It involves the M\"obius function. Indeed, if we apply the inverse morphism $\beta$ on the diagonal groupoid arrows $1_e$, we find:
\begin{equation}
\beta([1_e]) = \sum_{n \le e} \mu (n,e) n \, .
\end{equation}
To obtain the one-point function on the monoid,  we set the one-point function of these $|E(M)|$ elements equal to one while setting the ones of all non-diagonal arrows (pulled back to the monoid) to zero.
%\footnote{As a consequence, the one-point function can be non-zero on monoid elements that are not idempotents. Indeed, non-idempotents can have fixed points in a monoid, in contrast to the group case.} 
%

\subsubsection*{Remarks}
\begin{itemize}

\item As in the group case, we chose a particular theory. It reduces to the group theory special case when the monoid is a group. There are many other choices of one-point function allowed. There is one free parameter per (generalized) conjugacy class of the monoid, labeled by $(e,\mu)$.

\item We have implicitly worked with the irreducible representations of the monoid induced by the irreducible representations of the maximal groups $G_e$ associated to idempotents. It is a non-trivial task to decompose a representation of the monoid in terms of these induced irreducible representations. For groups, the decomposition can be determined by the character of the representation and its decomposition in terms of irreducible characters, easily determined using a standard inner product on class functions.
In fact, for monoids, the decomposition can be determined similarly, but the definition of the inner product invokes the M\"obius function of the partially ordered idempotents \cite{Steinberg06II}.

\item For a generic monoid, the open-closed theory is more difficult to formulate. One way to proceed would be to consider the monoid algebra divided out by its radical and to build the theory from this semisimple starting point. 

\item We note that the number of irreducible representations of an inverse monoid equals the number of its generalized conjugacy classes which are in bijection with the conjugacy classes of the maximal subgroups $G_e$ \cite{Steinberg}.

\item One can imagine further applications of these simple theories as universal subsectors of more complicated ones.
Indeed, partial symmetries generically lead to inverse semigroups (which are in our context a straightforward generalization of monoids)
\cite{Lawson} and there are applications of partial symmetries in quasicrystals \cite{Quasicrystals}, the factorization of operators on Hilbert spaces 
% Patterson book 
and integrable spin chains \cite{Padmanabhan:2017ekk}. Our topological theories capture combinatorial consequences of the presence of partial symmetries in two-dimensional systems with and without boundaries.

\end{itemize}

\section{Grand Canonical Open-Closed Symmetric Orbifolds}
\label{GrandCanonicalOpenClosed}
In this section, we reap a first reward of our analysis of inverse monoid topological quantum field theories. Recall that the closed $S_n$ topological quantum field theory describes principle $S_n$ bundles over Riemann surfaces. These correspond to ramified covers of surfaces of degree $n$.  In this section, we consider the monoid ${\cal P}_n$ that simultaneously captures covers of arbitrary degree up to $n$ \cite{Okounkov:2002cja,Benizri:2024mpx}. Using the tools developed in the previous section, we formulate an open-closed theory of arbitrary degree covers with boundaries. 
We first do this in a gauge variant manner with labeled sheets and then discuss the $S_n$  projection.

\subsection{The Finite Inverse Partial Permutation Monoid}
 Our monoid of choice is the  Ivanov-Kerov monoid ${\cal P}_n$ of partial permutations. The monoid has elements  
\begin{align}
(d,w) \in {\cal P}_n
\end{align}
where $d$ is a subset of $P_n=\{1,2,\dots,n \}$ and $w$ is a permutation of the elements of the set $d$. 
The multiplication of two partial permutations equals
\begin{align}
(d_1,w_1)(d_2,w_2) &= (d_1 \cup d_2, w_1 w_2)
\end{align}
where the permutations $w_1$ and $w_2$ are trivially extended over the union set $d_1 \cup d_2$. The unit is the empty set with no permutation. The inverse $(d,w)^\ast$ of a partial permutation is equal to $(d,w)^\ast=(d,w^{-1})$.\footnote{Note that there is another  notion of partial permutations. It  consists of bijections of a subset of $P_n$ to another subset and the composition is defined as the composition of functions (where it is defined). This monoid is called the symmetric inverse monoid $I_{P_n}$ and plays a role for finite monoids akin to the symmetric group for groups.
%(Wagner-Preston).
The dimension of the symmetric inverse monoid is $\sum_{k=0}^n \binom{n}{k}^2 k!$ while the dimension of the Ivanov-Kerov monoid is $\sum_{k=0}^n \binom{n}{k} k!$ (because for the first case, we choose $k$ elements twice). 
In this paper, unless explicitly stated otherwise, we  refer to the Ivanov-Kerov inverse monoid of partial permutations when referring to partial permutations. }
%The algebra $\mathbb{C}[{\cal P}_n]$ is semisimple. 
The semisimple algebra $\mathbb{C}[{\cal P}_n]$ governs the combinatorics of grand canonical symmetric orbifolds \cite{Benizri:2024mpx}. The center of the algebra has dimension $\sum_{k=0}^n \binom{n}{k} p(k)$ where $p(k)$ is the number of partitions of $k$ (as will become clear shortly).  We can consider the vector bundles on these points to determine the spectrum of  boundary conditions of the theory. We wish to provide the details of a particular choice of such theory, as discussed in generality in  section \ref{MonoidTQFT}.

\subsection{The Structure of  Partial Permutations}
\label{IKPartialPermutations}
Since the theory we constructed  depends on the  structure of the inverse monoid we determine the pertinent properties  
of the Ivanov-Kerov inverse monoid ${\cal P}_n$ of partial permutations. The idempotents of the partial permutation monoid are the identities on any subset $d$, namely $(d,1)$.\footnote{
Instead of identifying idempotents by the labels $(d,1)$ we will often use the label $d$ for brevity.} Idempotents are naturally partial ordered. By the composition of partial permutations, the partial ordering of the idempotents is (the reverse of) the one corresponding to subsets. We have $d' \le d$ if and only if $d \subseteq d'$. The maximum with respect to the partial order is the identity in the monoid, namely the empty set.  The partial order on the full monoid is such that  it is the subset structure  that is important (since we can grow the set $d$ of the partial permutation in any way we wish with an idempotent). Thus, we have $m \le n$ for partial permutations $m$ and $n$ if the permutation part agrees and $m$ has a larger (or equal) set on which it acts than does $n$. 
The ${\cal J}$ equivalence classes of the  monoid are determined by the union part of the partial permutation composition. This implies that monoid elements are equivalent if and only if they have the same set $d$ as their first member. Thus, we have $2^n$  ${\cal J}$ equivalence classes and one idempotent per ${\cal J}$ class.\footnote{For comparison, the  symmetric inverse monoid $I_{P_n}$ has $\binom{n}{k}$ idempotents per equivalence class of bijections on $k$ elements.}
The M\"obius function for the partial order of subsets is $\mu(d',d)=(-1)^{|d'|-|d|}$ for $ d' \subseteq d$. 
Thus,  the isomorphism of monoid and groupoid algebra 
\begin{equation}
\mathbb{C}[M_{IK}] \stackrel{\alpha}{\cong} \mathbb{C}[\text{Go}_{IK}] \cong \prod_{d \subset P_n} \mathbb{C} [S_d]
\end{equation}
becomes:
\begin{align}
\alpha ((d,w)) &= \sum_{d' \supseteq d} [(d',w)]
\nonumber \\
\alpha^{-1}([(d,w)]) = \beta ([(d,w)]) &= \sum_{d' \supseteq d} (-1)^{|d'|-|d|} (d',w) \, .
\end{align}
Note that the restriction that source and target must agree in order for two factors to multiply to a non-zero element in the groupoid effectively cuts up the partial permutation monoid into separate permutation groups. 
We further need the map from the groupoid algebra to the matrix algebra. This is straightforward because all multiplicities $n_d$ are one. Thus, the general results agree with the case analysis in \cite{IK}. 
The dimension of the algebra is:
\begin{equation}
\sum_{d \subset P_n} |d|! = \sum_{k=0}^n \binom{n}{k} k! \, \, .
\end{equation}
The center is the center of the symmetric group algebras and has dimension $\sum_{k=0}^n \binom{n}{k} p(k)$ as announced. While the center of the direct sum algebra is generated by the traditional group idempotents
(\ref{GroupIdempotent}), the center on the partial permutation side of the isomorphism has the mildly more complicated expression (\ref{MonoidIdempotent}): 
\begin{equation}
e_{d,\rho} = \frac{d_{|d|,\rho}}{S_{d}} \sum_{g \in S_d} \chi_\rho(g^{-1}) \sum_{d' \supset d} (-1)^{|d'|-|d|} (d',g) \, .
\end{equation}
It is understood that $g$ is trivially extended to the set $d'$ in the final partial permutation. 
The bulk one-point functions are
\begin{equation}
\langle e_{d,\rho} \rangle_{\text{bulk}} = \frac{d_{k,\rho}^2}{(k!)^2} \, .
\end{equation}

\subsubsection*{The Open Sector}
The irreducible representations of the monoid ${\cal P}_n$ are determined by those of the maximal symmetric subgroups $S_{d}$. The representations are inherited through the definition \cite{IK}:
\begin{equation}
\pi_{d',\lambda}((d,w) ) = \pi_\lambda (\phi_{d'}((d,w))) \quad \text{for} \quad d \subset d'
\quad \text{else} \quad 0
\end{equation}
where $\phi_{d'}$ is the projection onto the group subalgebra associated to the set  $d'$ (where we extend $w$ to $d'$) and the tableau of $\lambda$ has $|d'|$ boxes. This provides a complete list of inequivalent irreducible representations.   
The representation theory is semisimple and the homomorphisms and matrix algebra are treated as in the generic case. 

\subsection{The Permutation Invariant Subsector}

A way to  view the theories we constructed is as the twisted sectors of (a grand canonical version of) the topological symmetric orbifold of a trivial theory \cite{Benizri:2024mpx}. From that viewpoint, we wish to project onto the observables that are permutation invariant such that the action of the $S_n$ symmetric group can be interpreted as a gauge symmetry. Another argument to implement this projection is that we wish to make the theory independent of the particular labels $d$ we use. 

In order to project onto permutation invariants, 
we must identify the $S_n$ action on all the quantities we defined above. The action is induced by the action of the symmetric group $S_n$ on the label set $P_n$. In other words, we rename the labels inside the subsets $d \subset P_n$ throughout the whole construction. 
  We have an action of an element $\sigma$ of the symmetric group $S_n$ on the partial permutation  monoid which is \cite{IK}
\begin{equation}
(d,w) \mapsto (\sigma \cdot d,\sigma w \sigma^{-1}) \, .
\end{equation}
One can then define the invariant subalgebra ${\cal A}_n
= \mathbb{C}[{\cal P}_n]^{S_n}$ which 
lies in the center of the monoid algebra. 
The action of $S_n$ on ${\cal P}_n$ divides it into orbits $A_{\rho;n}$ labeled by partial partitions. These span the invariant subalgebra.  
The invariant subalgebra is again semisimple because it is a direct summand of a semisimple algebra $\mathbb{C}[{\cal P}_n]$ (which can be split into kernel and image of the projection operator).  
The number of orbits or the number of points in the spectrum of the algebra equals $\sum_{k=0}^n p(k)$.
The orbits $A_{\rho;n}$ form a linear basis of the algebra of invariants ${\cal A}_n$ \cite{IK}.
To construct the open theory, we exploit  a natural $S_n$ action on the representations of the algebra $\mathbb{C}[{\cal P}_n]$. A representation is associated to a fixed subset $d'$. Clearly, we will act with $\sigma \in S_n$ on the subset $d'$. We have the symmetric group action:
\begin{align}
\sigma \cdot \pi_{\lambda,d'} ((d,w)) &= \pi_\lambda ( \varphi_{\sigma d'} ((\sigma d,\sigma w \sigma^{-1}))) \, .
\end{align}
This is the  equivalence relation by which we mod out the set of boundary conditions.  The number of inequivalent irreducible representations becomes $\sum_{k=0}^n p(k)$,
as expected. 

Finally, we provide a few more remarks on the whole section.
\begin{itemize}
\item Previously, we noted that we have a choice of linear form which depends on the labels $(d,\lambda)$. 
 If we want to keep the symmetric group $S_n$ symmetry, we must choose a linear form which depends only on the cardinal number of $|d|$. An example weight is an exponential of an area to the power of the number of covering sheets. This implements an area deformation of the topological theory. See \cite{Benizri:2024mpx}.
 \item 
 In \cite{IK} it was demonstrated that one can take the $n$ to infinity limit. It exploits projections $\theta_m$
\begin{equation}
\theta_m: \mathbb{C}[{\cal P}_n] \rightarrow \mathbb{C}[{\cal P}_{m \le n}]
\end{equation}
which map the permutations onto partial permutations of the smaller set $P_m$. One can then take the projective limit \cite{IK} for $n \rightarrow \infty$.
 The infinite group of finite permutations of the positive integers acts on these algebras. Our remark is brief: we can similarly project the representation theory to find a limiting open-closed theory at infinite $n$.

\item It is important to note that both the grand canonical and the microcanonical perspectives have their merits. The microcanonical view on symmetric orbifolds is simple and standard and has broad applications in conformal field theory and statistical physics. The grand canonical perspective is convenient for, for instance, proving $n$-(in)dependence of certain structure constants \cite{IK,Li:2020zwo,Ashok:2023kkd} and for matching gravitational properties in a holographic duality \cite{Maldacena:2000hw,Giveon:1998ns,Eberhardt:2020bgq,Benizri:2024mpx,Benizri}. The relation between the two perspectives is provided, technically, by the M\"obius function of subsets.
%This is probably sensible.)

\end{itemize}

\section{Discussion and Conclusions}
\label{Conclusions}

A finite inverse monoid algebra over the complex numbers is semisimple. As such, the resulting bulk as well as boundary topological quantum field theory is standard. Nevertheless, as in the group theory case, even after this simple global perspective is attained many questions remain. For example, the closed theory with a finite gauge group has different bases and a difficult combinatorial  problem is to compute the structure constants of the algebra in the conjugacy class sum basis. Also, requiring a  geometric interpretation in terms of fiber bundles forces a particular choice of Frobenius linear form on the algebra. Some Frobenius algebra choices are more canonical, geometric or algebraic than others. We detailed one of these choices in a group theory language, while exhibiting the connection to the abstract perspective. 

For the finite inverse monoid algebra, we reached the same level of understanding. To obtain explicit formulas, it is necessary to precisely understand why the monoid algebra is semisimple which in turn requires detailing the isomorphism to matrix algebras. We identified two useful isomorphisms from the structure theory of inverse monoids as well as the relevant idempotents in the monoid algebra. The M\"obius function of the partially ordered set of idempotents played a crucial combinatorial role. Mastering these tools allows for the direct verification of the Cardy condition in the open-closed monoid theory. 

A first and motivating application is the formulation of a finite $n$  grand canonical theory of topological symmetric orbifolds with boundaries. We  understood better the isomorphy between grand canonical and microcanonical symmetric orbifolds and the partially ordered set combinatorics necessary to switch between these ensembles.
There may be  many more applications of the theories we constructed, in particular in contexts where partial symmetries have been identified to play an important role.

We conclude with a brief enumeration of related  research questions.
An interesting open problem is to match the lattice construction for the groupoid theory to a theory of principal bundles of groupoids. 
There may also be  a  non-commutative open string product formula analogue of the second quantized topological symmetric orbifold product \cite{Kaufmann:2002nx}.
It would be instructive to incorporate non-trivial homology of the monoid as in Dijkgraaf-Witten theory \cite{Dijkgraaf:1989pz}.
We have developed our formalism mainly for topological theories but a lot will carry over to physical grand canonical symmetric orbifold conformal field theories.
It will be rewarding to match combinatorial data in the presence of boundaries with supersymmetric correlators in holographic duals (as has been done for the conjugacy class algebra in the bulk). 

Finally, it would be beneficial to describe the open/closed Hurwitz/Gromov-Witten correspondence in detail.
We recall that the  closed Hurwitz/Gromov-Witten correspondence \cite{Okounkov:2002cja} can be interpreted as a gauge/gravity duality \cite{Benizri:2024mpx}. We believe we have prepared the ground for extending this  simple and exact gauge/gravity duality to include boundaries. 

\section*{Acknowledgments}
It is a pleasure to thank my colleagues and Lior Benizri in particular for stimulating discussions on these and related topics. 

%\newpage

 \appendix

 \section{Idempotents}
 \label{IdempotencyAppendix}
A crucial role in this paper is played by idempotency relations in the group and monoid algebra. We review the idempotency calculation for finite groups and extend it to the case of inverse monoids.
\subsection{Group Idempotents}
For a finite group $G$, idempotents in the group algebra associated to each irreducible representation $\rho$ are:
\begin{equation}
e_\rho = \sum_{g \in G} \frac{d_\rho}{|G|} \chi_\rho(g^{-1}) g  \, .
\end{equation}
These are orthogonal idempotents because:
\begin{align}
e_\lambda e_\rho &= \sum_{g,h \in G} \frac{ d_\lambda d_\rho}{|G|^2}  \chi_{\lambda}(h^{-1}) \chi_\rho(g^{-1}) h g  
\nonumber \\
&= \sum_{g,h \in G} \frac{ d_\lambda d_\rho}{|G|^2}  \chi_{\lambda}(g h^{-1}) \chi_\rho(g^{-1}) h
\nonumber \\
&= \delta_{\lambda,\rho} \sum_{h \in G} \frac{ d_\lambda}{|G|}  \chi_{\lambda}(h^{-1})  h
\nonumber \\
&= \delta_{\lambda,\rho} e_\lambda
 \, .
\end{align}
We used a character identity that can be proven using Schur orthogonality for matrices of irreducible representations. 
\subsection{Inverse Monoid Idempotents}
For a finite inverse monoid, the idempotents in the monoid algebra associated to each irreducible representation of the maximal subgroups $G_e$ (one per $J$ equivalence class) are:
\begin{equation}
e_{e,\rho} = \sum_{e \in E(J)} \left( \frac{d_{e,\rho}}{|G_e|} \sum_{s \in G_e} \chi_\rho(s^{-1}) \sum_{t \le s} t \mu(t,s) \right) \, .
\end{equation}
We confirm their nature:
\begin{align}
e_{f,\lambda} e_{e,\rho} &= \sum_{e \in E(J_e),f \in E(J_f)} \sum_{m \in G_f,s \in G_e} \frac{ d_{f,\lambda} d_{e,\rho}}{|G_e||G_f|}  \chi_{\lambda} (m^{-1}) \chi_\rho(s^{-1}) \sum_{n \le m,t \le s} 
\mu(n,m) \mu(t,s) 
n t
\nonumber \\
&= \sum_{e,f,m,s} 
\frac{ d_{f,\lambda} d_{e,\rho}}{|G_e||G_f|}  \chi_{\lambda} (m^{-1}) \chi_\rho(s^{-1}) \sum_{ t' \le ms}  \mu(t',ms) 
 t'
\nonumber \\
&=\delta_{e,f} \sum_{e \in E(J_e)} \sum_{m,s \in G_e} \frac{ d_{e,\lambda}^2} {|G_e|^2}  \chi_{\lambda} (m^{-1}) \chi_\rho(s^{-1}) \sum_{ t' \le ms}  \mu(t',ms) 
 t'
 \nonumber \\
&=\delta_{e,f} \sum_{e \in E(J_e)} \sum_{m,s' \in G_e} \frac{ d_{e,\lambda}^2} {|G_e|^2}  \chi_{\lambda} (m^{-1}) \chi_\rho( (s')^{-1} m) \sum_{ t' \le s'}  \mu(t',s') 
 t'
\nonumber \\
&= \delta_{e,f} \delta_{\lambda,\rho} \sum_{s' \in G_e} \frac{ d_{e,\lambda}}{|G_e|}  \chi_{\rho}((s')^{-1})  \sum_{t' \le s'} \mu(t',s') t'
\nonumber \\
&= \delta_{e,f} \delta_{\lambda,\rho} e_{e,\rho}
 \, .
\end{align}
We  needed
that $\beta$ is an isomorphism. Indeed, that property implies:
\begin{align}
\beta([mn]) &= \sum_{k \le mn} \mu(k,mn) k
\nonumber \\
= \beta([m]) \beta([n]) 
&= \sum_{k \le m} \mu (k,m) k \sum_{l \le n} \mu (l,n) l \, .
\end{align}
% Moreover, inverse preserves partial order \cite{Steinberg06}. 
This equality justifies the second line. Since $m$ and $s$ are group elements and since they are multiplied in the second line, they must correspond to the same idempotent, which is a fact used in the third line.  The other steps are the group theory calculations also featuring in the previous subsection.


\begin{thebibliography}{99}


%\cite{Atiyah:1989vu,Dijkgraaf:1989pz}
\bibitem{Atiyah:1989vu}
M.~Atiyah,
``Topological quantum field theories,''
Inst. Hautes Etudes Sci. Publ. Math. \textbf{68} (1989), 175-186
doi:10.1007/BF02698547
%499 citations counted in INSPIRE as of 05 Aug 2025

%\cite{Dijkgraaf:1989pz}
\bibitem{Dijkgraaf:1989pz}
R.~Dijkgraaf and E.~Witten,
``Topological Gauge Theories and Group Cohomology,''
Commun. Math. Phys. \textbf{129} (1990), 393
doi:10.1007/BF02096988
%861 citations counted in INSPIRE as of 05 Aug 2025

%\cite{Fukuma:1993hy}
\bibitem{Fukuma:1993hy}
M.~Fukuma, S.~Hosono and H.~Kawai,
``Lattice topological field theory in two-dimensions,''
Commun. Math. Phys. \textbf{161} (1994), 157-176
doi:10.1007/BF02099416
[arXiv:hep-th/9212154 [hep-th]].
%126 citations counted in INSPIRE as of 05 Aug 2025

\bibitem{Hurwitz}
A. Hurwitz, \"Uber die Anzahl der Riemann’schen Fl\"achen mit gegebenen 
Verzweigungspunkten, Math. Ann. {\bf 55} (1902) 53


%\cite{Mednyh,Atiyah:1989vu,Dijkgraaf:1989pz}
\bibitem{Mednyh}
A.D. Mednyh, Hurwitz problem on the number of nonequivalent coverings of a compact Riemann surface
Sib.Math.J. 23 (1982) 415 doi: 10.1007/BF00973499

%\cite{Maldacena:2000hw,Giveon:1998ns,Eberhardt:2020bgq}
\bibitem{Maldacena:2000hw}
J.~M.~Maldacena and H.~Ooguri,
``Strings in AdS(3) and SL(2,R) WZW model 1.: The Spectrum,''
J. Math. Phys. \textbf{42} (2001), 2929-2960
doi:10.1063/1.1377273
[arXiv:hep-th/0001053 [hep-th]].
%652 citations counted in INSPIRE as of 05 Aug 2025

%\cite{Giveon:1998ns,Eberhardt:2020bgq}
\bibitem{Giveon:1998ns}
A.~Giveon, D.~Kutasov and N.~Seiberg,
``Comments on string theory on AdS(3),''
Adv. Theor. Math. Phys. \textbf{2} (1998), 733-782
doi:10.4310/ATMP.1998.v2.n4.a3
[arXiv:hep-th/9806194 [hep-th]].
%514 citations counted in INSPIRE as of 05 Aug 2025

%\cite{Eberhardt:2020bgq}
\bibitem{Eberhardt:2020bgq}
L.~Eberhardt,
``Partition functions of the tensionless string,''
JHEP \textbf{03} (2021), 176
doi:10.1007/JHEP03(2021)176
[arXiv:2008.07533 [hep-th]].
%91 citations counted in INSPIRE as of 05 Aug 2025

%\cite{Okounkov:2002cja,Benizri:2024mpx}
\bibitem{Okounkov:2002cja}
A.~Okounkov and R.~Pandharipande,
``Gromov-Witten theory, Hurwitz theory, and completed cycles,''
Ann. Math. \textbf{163} (2006), 517-560
doi:10.4007/annals.2006.163.517
[arXiv:math/0204305 [math]].
%126 citations counted in INSPIRE as of 05 Aug 2025

%\cite{Benizri:2024mpx}
\bibitem{Benizri:2024mpx}
L.~Benizri and J.~Troost,
``Symmetric group gauge theories and simple gauge/string dualities,''
J. Phys. A \textbf{57} (2024) no.50, 505401
doi:10.1088/1751-8121/ad92ce
[arXiv:2404.12543 [hep-th]].
%2 citations counted in INSPIRE as of 05 Aug 2025

%\cite{Gross:1993hu,Cordes:1994fc,Benizri:2025xmz}
\bibitem{Gross:1993hu}
D.~J.~Gross and W.~Taylor,
``Two-dimensional QCD is a string theory,''
Nucl. Phys. B \textbf{400} (1993), 181-208
doi:10.1016/0550-3213(93)90403-C
[arXiv:hep-th/9301068 [hep-th]].
%438 citations counted in INSPIRE as of 05 Aug 2025

%\cite{Cordes:1994fc,Benizri:2025xmz}
\bibitem{Cordes:1994fc}
S.~Cordes, G.~W.~Moore and S.~Ramgoolam,
``Lectures on 2-d Yang-Mills theory, equivariant cohomology and topological field theories,''
Nucl. Phys. B Proc. Suppl. \textbf{41} (1995), 184-244
doi:10.1016/0920-5632(95)00434-B
[arXiv:hep-th/9411210 [hep-th]].
%376 citations counted in INSPIRE as of 05 Aug 2025

%\cite{Benizri:2025xmz}
\bibitem{Benizri:2025xmz}
L.~Benizri and J.~Troost,
``The String Dual to Two-dimensional Yang-Mills Theory Revisited,''
[arXiv:2502.02662 [hep-th]].
%2 citations counted in INSPIRE as of 05 Aug 2025

\bibitem{IK}
V.~Ivanov, S.~Kerov, ``The algebra of conjugacy classes in symmetric groups and partial
permutations,'' Journal of Mathematical Sciences, {\bf 107} (5), 4212-4230 (2001).




\bibitem{Steinberg}
B.~Steinberg, ``Representation theory of finite monoids,''  Springer, 2016.


%\cite{Lauda:2006mn}
\bibitem{Lauda:2006mn}
A.~D.~Lauda and H.~Pfeiffer,
``State sum construction of two-dimensional open-closed topological quantum field theories,''
J. Knot Theor. Ramifications \textbf{16} (2007), 1121-1163
doi:10.1142/S0218216507005725
[arXiv:math/0602047 [math.QA]].
%42 citations counted in INSPIRE as of 23 Jul 2025





%\cite{Belin:2021nck,Gaberdiel:2021kkp}
\bibitem{Belin:2021nck}
A.~Belin, S.~Biswas and J.~Sully,
``The spectrum of boundary states in symmetric orbifolds,''
JHEP \textbf{01} (2022), 123
doi:10.1007/JHEP01(2022)123
[arXiv:2110.05491 [hep-th]].
%27 citations counted in INSPIRE as of 05 Aug 2025

%\cite{Gaberdiel:2021kkp}
\bibitem{Gaberdiel:2021kkp}
M.~R.~Gaberdiel, B.~Knighton and J.~Vo{\v{s}}mera,
``D-branes in AdS$_{3}$ x S$^{3}$ x {\ensuremath{\mathbb{T}}}$^{4}$ at k = 1 and their holographic duals,''
JHEP \textbf{12} (2021), 149
doi:10.1007/JHEP12(2021)149
[arXiv:2110.05509 [hep-th]].
%45 citations counted in INSPIRE as of 05 Aug 2025

\bibitem{AN}
A.~Alekseevskii and S.~Natanzon,
``The algebra of bipartite graphs and Hurwitz numbers of seamed surfaces,''
 Izv. Math.  {\bf 72} (2008) 627


%\cite{Gopakumar:2011ev}
\bibitem{Gopakumar:2011ev}
R.~Gopakumar,
``What is the Simplest Gauge-String Duality?,''
[arXiv:1104.2386 [hep-th]].
%48 citations counted in INSPIRE as of 05 Aug 2025


%\cite{Li:2020zwo}
\bibitem{Li:2020zwo}
S.~Li and J.~Troost,
``The Topological Symmetric Orbifold,''
JHEP \textbf{10} (2020), 201
doi:10.1007/JHEP10(2020)201
[arXiv:2006.09346 [hep-th]].
%9 citations counted in INSPIRE as of 10 Sep 2025

%\cite{Elitzur:2011ug,Buryak:2020pgl}
\bibitem{Elitzur:2011ug}
S.~Elitzur, Y.~Oz, E.~Rabinovici and J.~Walcher,
``Open/Closed Topological CP1 Sigma Model Revisited,''
JHEP \textbf{01} (2012), 101
doi:10.1007/JHEP01(2012)101
[arXiv:1106.2967 [hep-th]].
%0 citations counted in INSPIRE as of 05 Aug 2025

%\cite{Buryak:2020pgl}
\bibitem{Buryak:2020pgl}
A.~Buryak, A.~N.~Zernik, R.~Pandharipande and R.~J.~Tessler,
``Open CP1 descendent theory I: The stationary sector,''
Adv. Math. \textbf{401} (2022), 108249
doi:10.1016/j.aim.2022.108249
[arXiv:2003.00550 [math.SG]].
%1 citations counted in INSPIRE as of 05 Aug 2025

%\cite{Moore:2006dw}
\bibitem{Moore:2006dw}
G.~W.~Moore and G.~Segal,
``D-branes and K-theory in 2D topological field theory,''
[arXiv:hep-th/0609042 [hep-th]].
%169 citations counted in INSPIRE as of 05 Aug 2025

\bibitem{Moore}
G.~Moore, ``A Few Remarks on Topological Field Theory,''
downloaded from the URL
https://www.physics.rutgers.edu/\~\,gmoore/695Fall2015/TopologicalFieldTheory.pdf on 5 August, 2025. 

\bibitem{Pennig}
U.~Pennig, ``Gauge Theory with Finite Gauge Groups,''
downloaded from the URL
https://upennig.weebly.com/uploads/7/4/0/3/74037187/2d-tqft.pdf
on 5 August, 2025.


%\cite{Lazaroiu:2000rk}
\bibitem{Lazaroiu:2000rk}
C.~I.~Lazaroiu,
``On the structure of open - closed topological field theory in two-dimensions,''
Nucl. Phys. B \textbf{603} (2001), 497-530
doi:10.1016/S0550-3213(01)00135-3
[arXiv:hep-th/0010269 [hep-th]].
%122 citations counted in INSPIRE as of 29 Aug 2025

\bibitem{Koch}
J.~Koch,``Frobenius algebras and 2D topological quantum field theory,'' London Mathematical Society Student Texts {\bf 59} (2004).


%\cite{Lawson,Quasicrystals,Padmanabhan:2017ekk}
\bibitem{Lawson}
M.~Lawson,``Inverse semigroups: the theory of partial symmetries,'' World Scientific, 1998. % Brownian motion, automata.

\bibitem{Steinberg06}
B.~Steinberg, ``Möbius functions and semigroup representation theory,'' Journal of Combinatorial Theory, Series A {\bf 113.5} (2006), 866-881.

\bibitem{Steinberg06II} 
B.~Steinberg, ``Möbius functions and semigroup representation theory II: Character formulas and multiplicities,'' Advances in Mathematics,  {\bf 217.4} (2008), 1521-1557.




\bibitem{Quasicrystals}
 D.P. Di Vincenzo and P.J. Steinhardt, Quasicrystals: The State of the Art, World Scientific,
Singapore 1991.


%\cite{Padmanabhan:2017ekk}
\bibitem{Padmanabhan:2017ekk}
P.~Padmanabhan, S.~J.~Rey, D.~Teixeira and D.~Trancanelli,
``Supersymmetric many-body systems from partial symmetries {\textemdash} integrability, localization and scrambling,''
JHEP \textbf{05} (2017), 136
doi:10.1007/JHEP05(2017)136
[arXiv:1702.02091 [hep-th]].
%16 citations counted in INSPIRE as of 05 Aug 2025

%\cite{Ashok:2023kkd}
\bibitem{Ashok:2023kkd}
S.~K.~Ashok and J.~Troost,
``The operator rings of topological symmetric orbifolds and their large N limit,''
JHEP \textbf{04} (2024), 039
doi:10.1007/JHEP04(2024)039
[arXiv:2309.17052 [hep-th]].
%5 citations counted in INSPIRE as of 19 Sep 2025

\bibitem{Benizri}
L.~Benizri, ``
Grand-Canonical Symmetric Orbifold Theories'', to appear.

%\cite{Kaufmann:2002nx}
\bibitem{Kaufmann:2002nx}
R.~M.~Kaufmann,
``Second quantized Frobenius algebras,''
Commun. Math. Phys. \textbf{248} (2004), 33-83
doi:10.1007/s00220-004-1090-y
[arXiv:math/0206137 [math.AG]].
%7 citations counted in INSPIRE as of 17 Sep 2025


\end{thebibliography}
\end{document}